\documentclass{jfm}
\usepackage{graphicx}
\usepackage{epstopdf, epsfig}
\usepackage{amsmath}
\usepackage{subcaption}
\usepackage{placeins}
\usepackage{comment}
\def\munch{M\"unch}

\shorttitle{Taylor Dispersion in Osmotically Driven Laminar Flow in Phloem}
\shortauthor{M. Nakad, T. Witelski, J.C. Domec, S. Sevanto, and G. Katul}

\title{Taylor Dispersion in Osmotically Driven Laminar Flows in Phloem}

\author{M. Nakad\aff{1}
  \corresp{\email{mazen.nakad@duke.edu}},
  T. Witelski\aff{2},
  J.C. Domec\aff{1,3},
  S. Sevanto\aff{4},
  G. Katul\aff{1,5}}

\affiliation{\aff{1} Nicholas School of the Environment, Duke University, Durham, NC, USA \\
\aff{2} Department of Mathematics, Duke University, Durham, NC, USA
\\
\aff{3} Bordeaux Sciences Agro, UMR 1391 INRA-ISPA, France
\\
\aff{4} Earth and Environmental Sciences Division, Los Alamos National Laboratory, Los Alamos, NM, USA
\\
\aff{5} Department of Civil and Environmental Engineering, Duke University, Durham, NC, USA}

\begin{document}
\maketitle
\begin{abstract}
Sucrose is among the main products of photosynthesis that are deemed necessary for plant growth and survival. It is produced in the mesophyll cells of leaves and translocated to different parts of the plant through the phloem. Progress in understanding this transport mechanism remains fraught with experimental difficulties thereby prompting interest in theoretical approaches and laboratory studies. The \munch\ pressure and mass flow model is one of the commonly accepted hypotheses for describing the physics of sucrose transport in the phloem systems. It is based on osmosis to build an energy potential difference between the source and the sink. The flow responding to this energy potential is assumed laminar and described by the Hagen-Poiseulle equation. This study revisits such osmotically driven flow in tubes by including the effects of Taylor dispersion on mass transport, which has not been considered in the context of phloem flow. Taylor dispersion is an effect in tube flow where shear flow can increase the effective transport of species. It is shown that in addition to the conventional Taylor dispersion diffusive correction derived for closed pipe walls, a new adjustments to the mean advective terms, arise because of osmotic effects.  These new terms act as local sources and sinks of sucrose, though their overall average effect is zero.  Because the molecular Schmidt number is very large for sucrose in water, the role of a radial P\'eclet number emerges as controlling the sucrose front speed and travel times above and beyond the much studied \munch\ number.  This study establishes upper limits on expected Taylor dispersion speed-up of sucrose transport.
\end{abstract}


\section{Introduction}
\begin{figure}
    \begin{subfigure}[t]{0.5\textwidth}
        \centering
        \includegraphics[width = \textwidth]{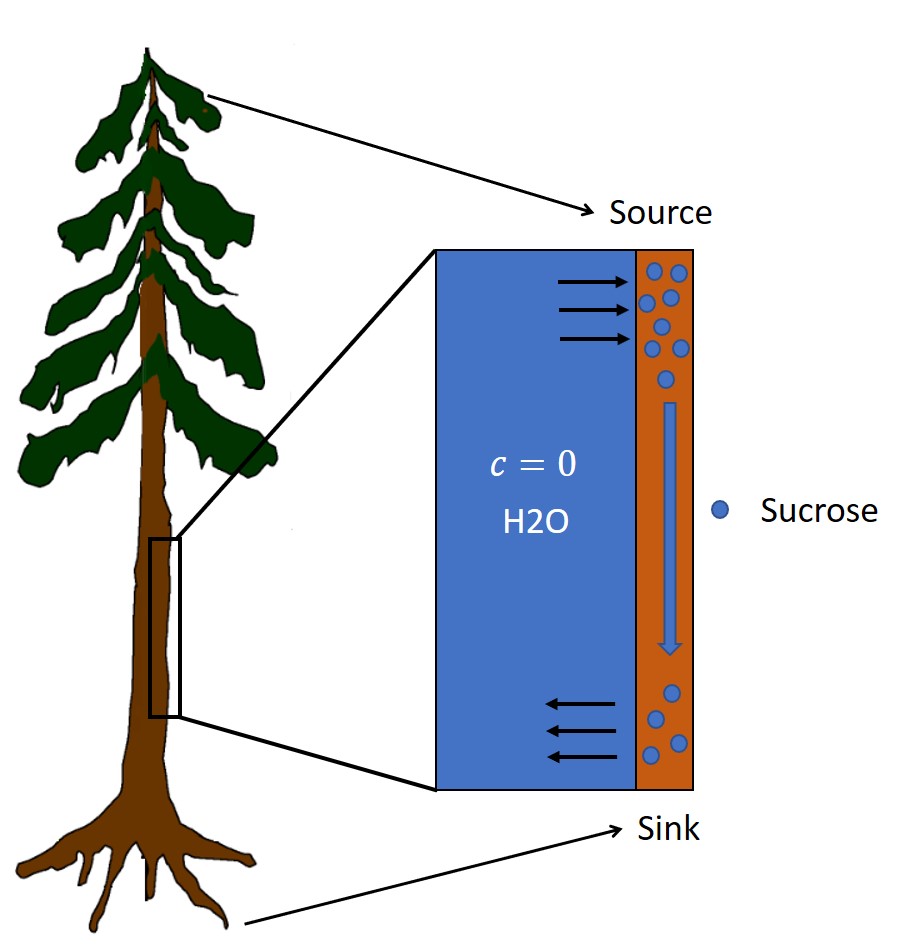}
        \caption{(\textit{a})}
        \label{fig:problem_plant}
    \end{subfigure}
    \begin{subfigure}[t]{0.5\textwidth}
        \centering
        \includegraphics[width = \textwidth]{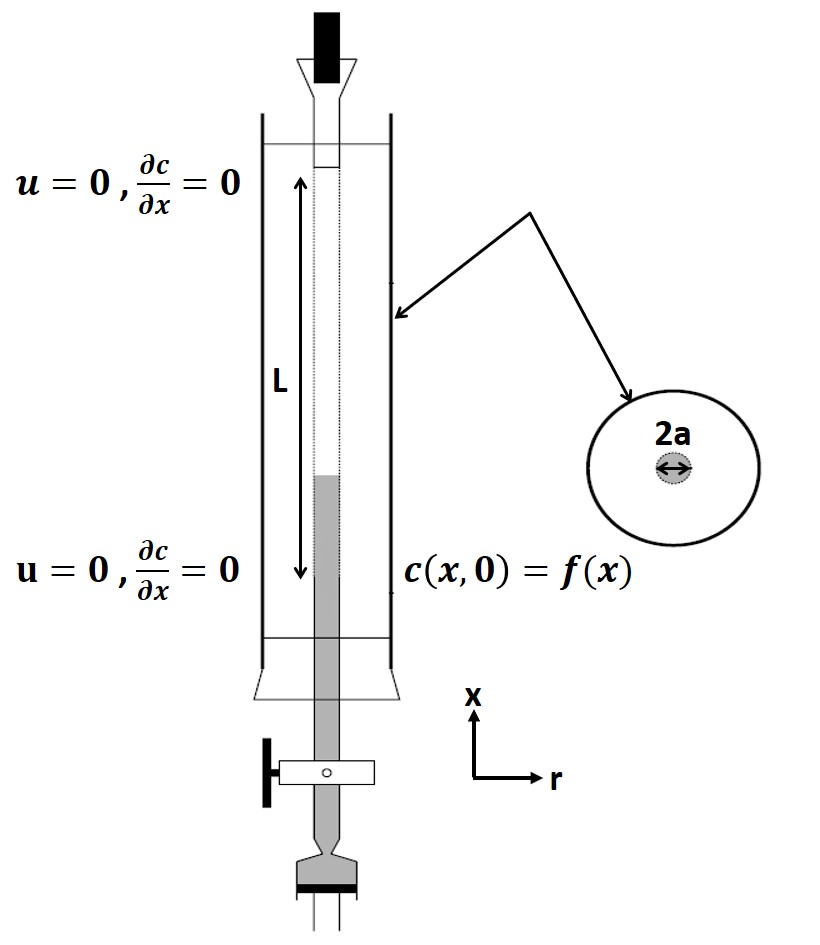}
        \caption{(\textit{b})}
        \label{fig:problem_experiment}
    \end{subfigure}
    \caption{(\textit{a}) Water and sucrose transport in the phloem (orange) driven by water inflow from and back into the xylem (blue) from source (i.e. leafs) to sink (i.e. roots). (\textit{b}) Schematic of the experiment used by \citet{jensen2009osmotically} to describe sucrose transport (side and top views). The phloem geometry is assumed to be a long and narrow tube of length $L$ and radius $a$. The xylem is assumed to be a naturally  filtered  water reservoir that covers the semi-permeable tube all around. Sucrose is loaded into the tube at $t = 0$ in a radially uniform manner and axially described by a smooth function $f(x)$.  Sucrose is conserved in the tube during the entire experiment. The axial and radial coordinates as well as the boundary conditions used are shown.}
    \label{fig:problem}
\end{figure}

 The physics of sucrose transport in plants, introduced in the early 1930s by the forestry scientist E. \munch\ \citep{munch1930stoffbewegungen}, continues to be the workhorse model today though this hypothesis is still not free from controversies \citep{curtis1933comparison,spanner1958translocation,christy1973mathematical,fensom1981problems,lang1983turgor,thompson2003application,minchin2005new,ryan2014phloem,savage2017maintenance,knoblauch2017actually,sevanto2018drought,huang2018transport}. The \munch\ hypothesis assumes that sucrose molecules produced during leaf photosynthesis in mesophyll cells are loaded into phloem tubes (figure \ref{fig:problem_plant}).  Through osmosis, water is then pulled into the phloem from adjacent cells, or xylem vessels, creating a positive pressure that pushes water along the phloem tube towards sink tissues where sucrose is consumed or converted to other forms for storage (figure \ref{fig:problem_plant}).  Because the sucrose concentration in these sink tissues is much smaller than in source tissues, the driving force for water movement in the phloem system can then be established.  The elegance, plausibility, simplicity, and partial experimental support 'endowed' this hypothesis with broad acceptance in plant physiology \citep{wardlaw1974phloem,housley1977estimation,rand1983fluid,van2003phloem,pickard2009simplest,mencuccini2010significance,jensen2011optimality,knoblauch2012structure,nikinmaa2013assimilate,knoblauch2016testing,jensen2016sap,jensen2018phloem,konrad2018xylem}.
\par

The main critique of the \munch\ hypothesis, which continues to draw research interest even today, is whether such a driving force allows sucrose to be loaded and transported sufficiently fast over long distances as may be expected in tall trees \citep{fensom1981problems,turgeon2010puzzle}.  Best estimates of sucrose concentrations in leaves raises some concerns about the generality and utility of the \munch\ hypothesis.  It has been reported that sucrose concentration in some leaves of tall trees is smaller than in shorter vegetation \citep{fensom1981problems,turgeon2010puzzle}.  Such concentration contrasts are not compatible with calculations that require the effective hydraulic conductivity to diminish with increased tube length \citep{knoblauch2016testing,knoblauch2017actually,savage2017maintenance} assumed here to be proportional to plant height.
\par
The focus here is on Taylor dispersion, an overlooked mechanism that enhances the effective hydraulic conductivity of tubes, such as phloem conduits \citep{taylor1953dispersion}.  Taylor dispersion can increase mass transport well beyond molecular diffusion in closed wall tubes (that are used in the calculations of phloem effective hydraulic conductivity).  However, Taylor dispersion in osmotically driven flow such as the one described by the \munch\ hypothesis (i.e. tube walls are membrane) leads to further adjustments  apart from the apparent increase in diffusion. In this study we uncover the physics of those adjustments and quantify their magnitude and effects on phloem flow and sucrose transport rates. 
 
\section{Theory}
\label{sec:theory}
The basic equations describing sucrose transport in plants are first reviewed. Since the focus is on scalar mass transport mechanisms, the conductive phloem geometry is simplified to permit analytical foresight (figure \ref{fig:problem_experiment}).  It is approximated by a long tube with length $\textit{L}$ and radius $\textit{a}$ connecting sucrose production at the leaf with sinks in the root (figure \ref{fig:problem}).  The phloem sieve tubes are long and narrow meaning that any dynamic scaling on flow variables will be subject to the slender geometry with aspect ratio $\epsilon = a / L\ll1$. 
\par
The tube surface area is covered by a membrane with uniform permeability $\textit{k}$ that allows water molecules, but not sucrose molecules, to be exchanged with the surrounding aqueous environment. Because the tube is effectively immersed in a water reservoir, the treatment of water flow can be achieved by placing the tube in a vertical position so that $x$ defines the longitudinal direction and $r$ defines the radial direction from the center of the tube (figure \ref{fig:problem_experiment}). Sucrose molecules enter the bottom of the tube at $x=0$ and propagate within the tube until $x=L$.  The tube is closed at $x=L$.
\par 
Throughout, sucrose concentration is denoted as $c$, fluid pressure as $p$, longitudinal velocity component as $u$, and radial velocity component as $v$.  The $u$ in plants is small, and therefore the flow can be approximated as a low Reynolds number flow (values  from 20 to 500 in the region of laminar flow), where the Reynolds number is defined as $\Rey=2a u/\nu\ll 1$, $\nu=\mu/\rho$ is the kinematic viscosity, $\mu$ and $\rho$ are the dynamic viscosity and density of water, respectively.  Hence, inertial effects in the longitudinal momentum balance can be ignored relative to viscous stresses.  Frictional losses due to the presence of sieve plates within the phloem are also momentarily ignored though in some cases, this loss can be significant \citep{knoblauch2016testing}.  This setup does not reproduce all the geometric complexities of the phloem network in plants. The simplifications here are intended as a logical starting point for exploring Taylor dispersion in osmotically driven flow in idealized settings.
\par
For mass transport, the molecular diffusion coefficient of sucrose $D=\nu/Sc$, where $Sc$ is the molecular Schmidt number ($>10,000$ for sucrose in water), is small.  The fact that sucrose transport occurs at such a high $Sc$ implies that the advective transport term in the scalar mass balance equation cannot be ignored (unlike in the momentum balance).  The strength of scalar advective to diffusive contributions are quantified using the P\'eclet number $\Pen=2a u/D$, which can also be expressed as $\Pen = \Rey \cdot Sc$. While $\Rey\ll 1$, the advective transport in the scalar mass balance equation need not be small precisely because $Sc\gg 1$.
\par
In the following sections, the system of equations that describe the physics of sucrose transport in plants is presented. The governing equations and their associated assumptions are first discussed. Next, the derivation of Taylor dispersion in osmotically driven flows are featured after area-averaging the governing equations. A brief description of the so-called \munch\ mechanism, which have been derived and reviewed elsewhere \citep{jensen2009osmotically,jensen2016sap}, is then presented while accommodating Taylor dispersion. Finally, a scaling analysis is used to demonstrate the existence of two distinct flow regimes based on the magnitude of the \munch\ number, which is defined as the ratio of the axial (mainly viscous) to membrane flow resistance \citep{jensen2009osmotically,jensen2016sap}. The focus of the results and discussion is on the consequences of Taylor dispersion within these two 'end-member' flow regimes.

\subsection{The governing equations}
\label{sec:point_equations}
It is assumed that water is an incompressible Newtonian fluid, with density $\rho$ and viscosity $\mu$, satisfying the continuity equation
\begin{equation}
    \frac{1}{r}\frac{\p (r v)}{\p r} + \frac{\p u}{\p x} = 0.
    \label{eq:continuity}
\end{equation}
For very high $c$, $\rho$ and $\mu$ need not be constant and can vary with $c$.  However, for low $c$, this dependency can be ignored.  Assuming cylindrical symmetry, the flow of water in the tube is described by momentum balance equations in the axial and radial directions
\begin{equation}
\begin{aligned}
    \rho \left( \frac{\p u}{\p t}  + u \frac{\p u}{\p x} + v \frac{\p u}{\p r} \right) = - \frac{\p p}{\p x} + \mu \left[\frac{\p^2 u}{\p x^2} + \frac{1}{r}\frac{\p}{\p r}\left(r \frac{\p u}{\p r}\right)\right], \\
    \rho \left( \frac{\p v}{\p t} + u \frac{\p v}{\p x} + v \frac{\p v}{\p r} \right) = - \frac{\p p}{\p r} + \mu \left[\frac{\p^2 v}{\p x^2} + \frac{1}{r}\frac{\p}{\p r}\left(r \frac{\p v}{\p r}\right) - \frac{v}{r^2}\right].
    \label{eq:NSE}
\end{aligned}
\end{equation}
Equations (\ref{eq:NSE}) assume that there are no external forces on the fluid and that the gravitational forces are negligible \citep{thompson2003application}.  The normalized variables defined by $u = u_0 U$, $v = v_0 V$, $p = p_0 P$, $r = a R$ and $x = L X$ are introduced, where $u_0$, $v_0$, and $p_0$ are characteristic axial velocity, radial velocity, and pressure.  The characteristic length scales in the axial and radial dimensions are the tube length $L$ and radius $a$. The radial velocity scale $v_0 = \epsilon u_0$ is determined from the continuity equation, and $p_0 = (L \mu u_0) / a^2$ is the viscous pressure scale.
The nondimensional form for the fluid continuity is then
\begin{equation}
    \frac{1}{R}\frac{\p (R V)}{\p R} + \frac{\p U}{\p X} = 0,
    \label{eq:continuity_nondimensional}
\end{equation}
and the Navier-Stokes equations for the axial and radial velocities at steady state are
\begin{equation}
\begin{aligned}
    \Rey~ \epsilon \left(U \frac{\p U}{\p X} + V \frac{\p U}{\p R}\right) = - \frac{\p P}{\p X} + \epsilon^2\frac{\p^2 U}{\p X^2} + \frac{1}{R}\frac{\p}{\p R}\left(R \frac{\p U}{\p R}\right), \\
    \Rey~ \epsilon^2 \left(U \frac{\p V}{\p X} + V \frac{\p V}{\p R}\right) = - \frac{\p P}{\p R} + \epsilon^2 \left[\epsilon^2 \frac{\p^2 V}{\p X^2} + \frac{1}{R}\frac{\p}{\p R}\left(R \frac{\p V}{\p R}\right) - \frac{V}{R^2}\right].
    \label{eq:NSE_nondimensional}
\end{aligned}
\end{equation}
As in lubrication theory, when the reduced Reynolds number tends to zero (i.e. $\epsilon~\Rey \ttz$), the leading order terms in equation (\ref{eq:NSE_nondimensional}) satisfy
\begin{equation}
    \frac{1}{R}\frac{\p}{\p R}\left(R \frac{\p U}{\p R}\right) = \frac{\p P}{\p X},\qquad
    \frac{\p P}{\p R} = 0.
    \label{eq:NSE_nondimensional_first_order}
\end{equation}
From equation (\ref{eq:NSE_nondimensional_first_order}), the leading order term of the axial and radial velocities can be obtained as a function of the pressure gradient inside the tube using the boundary conditions
\begin{equation}
  U(R = 1) = 0, \qquad
  \frac{\partial U}{\partial R}\bigg|_{R = 0} = 0, \qquad
  V(R = 0) = 0.
  \label{eq:BC_velocities}
\end{equation}
The first boundary condition in equation (\ref{eq:BC_velocities}) states that the axial velocity within the membrane is zero (though the radial velocity is finite at $R=1$ due to osmosis as later discussed). The second and third boundary conditions are derived from symmetry considerations at the center of the pipe.  Combining equations (\ref{eq:NSE_nondimensional_first_order}) with the continuity equation (\ref{eq:continuity_nondimensional}) and imposing the aforementioned boundary conditions (\ref{eq:BC_velocities}), the axial and radial velocities are given by
\begin{equation}
    U = - \frac{1}{4} \frac{\p P}{\p X} \left(1 - R^2 \right),\qquad 
    V = \frac{1}{4} \frac{\p^2 P}{\p X^2} \left(\frac{R}{2} - \frac{R^3}{4} \right).
    \label{eq:velocities_nondimensional}
\end{equation}
For completeness, these equations are also expressed in dimensional form as
\begin{equation}
    u = - \frac{1}{4 \mu} \frac{\p p}{\p x} \left(a^2 - r^2 \right),\qquad
    v = \frac{1}{4 \mu} \frac{\p^2 p}{\p x^2} \left(\frac{a^2}{2} r - \frac{r^3}{4} \right).
    \label{eq:velocities}
\end{equation}
Equation (\ref{eq:velocities}) represent the velocity components variation as first order approximation in the limit of small reduced Reynolds numbers.  While the axial velocity profile is virtually identical in mathematical form to the Hagen-Poiseuille (HP) equation derived for closed pipes, the radial velocity is not. In closed pipes, the no-slip condition at the pipe wall ($R=1$) and symmetry considerations at the center of the pipe ($R=0$) necessitate $v=0$ everywhere in the pipe, which is not the case here due to the presence of a membrane on the conduit walls.
\par
The use of the HP approximation to describe water movement in the phloem has been the subject of some debate \citep{henton2002revisiting,cabrita2013hydrodynamics,jensen2012analytic,thompson2003application,phillips1993asymptotic, jensen2009osmotically, thompson2003scaling,weir1981analysis}.
The main cause in this debate has been the assumption of an externally imposed constant pressure gradient ${\partial p}/{\partial x}$ routinely invoked in conventional derivation of the HP equation \citep{phillips1993asymptotic}.  A constant ${\partial p}/{\partial x}$ requires ${\partial^2 p}/{\partial x^2}=0$ and consequently $v=0$ everywhere (including at the boundary $r=a$). The expressions for $u$ and $v$ featured in equation (\ref{eq:velocities}) are straightforward and are compatible with the HP assumptions of a force-balance between pressure gradients and viscous stresses. In osmotically driven flow, the representation of the pressure and its gradients will be elaborated upon later.  While the combination of the continuity equation and the two momentum balance equations provide three equations in three unknowns ($p$, $v$, and $u$), they remain incomplete because an additional boundary condition on $v$ at $r=a$ is required. This boundary condition must be supplied by the membrane physics and osmoregulation. 
\par 
The conservation of scalar mass, which adds one more unknown and one more equation for $c$, is derived using the Reynolds transport theorem.  The sucrose mass fluxes in axial and radial directions are assumed to be a linear superposition of advection and molecular diffusion. The equation for the scalar mass balance can be expressed as
\begin{equation}
    \frac{\p c}{\p t} + \frac{\p (u c)}{\p x} + \frac{1}{r}\frac{\p (r v c)}{\p r} =
    D \frac{\p^2 c}{\p x^2} + D \frac{1}{r}\frac{\p}{\p r}\left(r \frac{\p c}{\p r} \right).
    \label{eq:advection_diffusion}
\end{equation}
The initial condition is that at $t = 0$, (and $L = 0$) sucrose enters the tube in a radially uniform manner. However, axially, an initial front $c = f(x)$ is prescribed that is the same as the one used by \citet{jensen2009osmotically} to ensure a smooth initial profile along the tube (figure \ref{fig:problem_experiment}). Along the axial direction, a no-mass flux at $x=L$ (the tube is closed at this end) and an externally specified uniform concentration at the source ($x=0$) suffice.  Along the radial direction, symmetry considerations provide one boundary condition as before, which is ${\partial c}/{\partial r}=0$ at $r=0$ and the no-mass flux at $r=a$ provide the second boundary condition where $v c - D \p c/\p r = 0$. However, as discussed before, another boundary condition is needed for $v$ at $r=a$ to mathematically close the problem. The equation providing closure to both $c$ and $v$ arises from the membrane physics. This equation describes the radial flow of water from the surrounding reservoir into the tube due to osmosis. This equation is best formulated as a boundary condition that relates the radial velocity $v$ to the driving gradient for water movement involving the fluid pressure $p$ and the osmotic potential (that varies with $c$) at $r = a$.  It is given as a Darcy type flow expression (described by \citet{iberall1973physics})
\begin{equation}
    v = k \left(p - \Pi_b\right),
    \label{eq:membrane_BC}
\end{equation}
where $k$ is the membrane permeability and $\Pi_b$ is the osmotic potential at the membrane ($r=a$ or $R=1$). For small $c$, the van't Hoff relation ($\Pi = R_g T c$) can be used to relate the osmotic potential $\Pi$ to the sucrose concentration at the membrane ($c=c_b$ at $r=a$), where $R_g$ is the gas constant and $T$ is the absolute temperature (assumed constant throughout).  The membrane physics provides the required boundary condition to link $v$ to $p$ and $c$ thereby providing the necessary closure to solving the 4 equations with 4 unknowns $c$, $u$, $v$, and $p$.  It is to be noted that at high $c$, not only the van't Hoff relation requires modification but $\rho$, $\nu$, and $D$ also become dependent on $c$.  This high sucrose concentration limit is outside the scope of the Taylor dispersion analysis featured next.

\subsection{Taylor Dispersion in Osmotically Driven Flows}
\label{sec:TD}
To elucidate the role of Taylor dispersion, equation (\ref{eq:advection_diffusion}) is averaged over the cross-sectional area of the tube. As a first approximation, two simplifications are made: (i) the axial velocity is much larger than the radial one (i.e. $\epsilon \ll 1$ and $v_0 = \epsilon  u_0$) while maintaining a finite $v_0$, and (ii) radial concentration changes are assumed to be small relative to their overall area-averaged value at any $x$. These two approximations remove the third term on the left hand side of equation (\ref{eq:advection_diffusion}) and the second term on the right hand side.  The second term on the left side of equation (\ref{eq:advection_diffusion}), area averaged $\p \overline{u c}/{\p x}$, is interesting given its connection to the original work of Taylor and is now explored. Even in closed pipes, this term is not $\p \bar{u} \bar{c}/{\p x}$. As noted by Taylor \citep{taylor1953dispersion}, the interaction between radial velocity and concentration variations adds an apparent diffusion term $D_d=a^2 \bar{u}^2 (48 D)^{-1} \p^2 \bar{c}/{\p x^2}$ that is labelled as dispersion.   Its effect is to monotonically increase the apparent diffusion coefficient $D+D_d$ and whence the name Taylor dispersion \citep{taylor1953dispersion}. In this section, a new Taylor dispersion term is derived and its effect for osmotically driven flows is discussed.  To arrive at expressions linking $\overline{u c}$ to $\bar{u} \bar{c}$, flow properties are decomposed into area-averaged and deviation components given as $c = \bar{c}(x,t) + \tilde{c}(x,r,t)$, $u = \bar{u}(x,t) + \tilde{u}(x,r,t)$ and $v = \bar{v}(x,t) + \tilde{v}(x,r,t)$, where
\begin{equation}
    \bar{c} = {2\over a^2}\int_0^a rc\,\mathrm{d}r,
\end{equation}
and similarly for other quantities the average of any deviation is identically zero.
\par
Inserting the decomposed variables into equation (\ref{eq:advection_diffusion}) leads to
\begin{equation}
\begin{split}
  \bar{c}_t + \tilde{c}_t + (\bar{c}\bar{u})_x + (\bar{c}\tilde{u})_x +
  (\tilde{c}\bar{u})_x + (\tilde{c}\tilde{u})_x \\
  + \frac{1}{r}(r\bar{c}\bar{v})_r + \frac{1}{r}(r\bar{c}\tilde{v})_r + \frac{1}{r}(r\tilde{c}\bar{v})_r + \frac{1}{r}(r\tilde{c}\tilde{v})_r \\
  = D\bar{c}_{xx} + D\tilde{c}_{xx}
  + D\frac{1}{r}(r\tilde{c}_r)_r + D\frac{1}{r}(r\bar{c}_r)_r,
\end{split}
  \label{eq:cm_decomposed}
\end{equation}
where differentiation is now written with respect to the subscripted variables. Averaging equation (\ref{eq:cm_decomposed}) radially while removing the last term on the right side of the equation ($\bar{c}$ only varies in $x$ and $t$), the area-averaged equation is
\begin{equation}
\begin{split}
    \bar{c}_t +
    \frac{2}{a^2}\int_0^a r\tilde{c}_t \mathrm{d} r + 
    \frac{2}{a^2}\int_0^a r\left(\bar{c}\bar{u}\right)_x \mathrm{d} r +
    \frac{2}{a^2}\int_0^a r\left(\bar{c}\tilde{u}\right)_x \mathrm{d} r \\ 
    + \frac{2}{a^2}\int_0^a r\left(\tilde{c}\bar{u}\right)_x \mathrm{d} r +
    \frac{2}{a^2}\int_0^a r\left(\tilde{c}\tilde{u}\right)_x \mathrm{d} r + 
    \frac{2}{a^2}\int_0^a \left(r\bar{c}\bar{v}\right)_r \mathrm{d} r \\
    + \frac{2}{a^2}\int_0^a \left(r\bar{c}\tilde{v}\right)_r \mathrm{d} r +
    \frac{2}{a^2}\int_0^a \left(r\tilde{c}\bar{v}\right)_r \mathrm{d} r +
    \frac{2}{a^2}\int_0^a \left(r\tilde{c}\tilde{v}\right)_r \mathrm{d} r  \\
  = \frac{2}{a^2}D\int_0^a \left(r\bar{c}_{xx}\right) \mathrm{d} r + 
  \frac{2}{a^2}D\int_0^a \left(r\tilde{c}_{xx}\right) \mathrm{d} r +  
  \frac{2}{a^2}D\int_0^a \left(r\tilde{c}_r\right)_r \mathrm{d} r.
  \label{eq:cm_decomposed_averaged}
\end{split}
\end{equation}
Eliminating terms that are the averages of deviations and evaluating other explicit integrals,  equation (\ref{eq:cm_decomposed_averaged}) becomes:
\begin{equation}
\begin{split}
    \bar{c}_t +
    (\bar{c}\bar{u})_x + \frac{2}{a^2}\int_0^a r\left(\tilde{c}\tilde{u}\right)_x \mathrm{d} r + \frac{2}{a}\left[c v - D \frac{\p \tilde{c}}{\p r}\right]_{r = a} 
    = D \bar{c}_{xx}.
    \label{eq:cm_averaged}
\end{split}
\end{equation}
The zero-mass flux boundary condition at the membrane, $v c - D\p c/\p r \vert_{r=a} = 0$, is enforced so that no sucrose molecules cross the membrane.  Hence, the area averaged equation satisfying this boundary condition (while noting that $\p c/\p r = \p \tilde{c}/\p r$) is
\begin{equation}
\begin{split}
  \bar{c}_t + (\bar{c}\bar{u})_x + \frac{2}{a^2}\int_0^a r\left(\tilde{c}\tilde{u}\right)_x \mathrm{d} r = D\bar{c}_{xx}.
\end{split}
  \label{eq:cm_averaged_final}
\end{equation}
 To determine $2{a^{-2}}\int_0^a r\left(\tilde{c}\tilde{u}\right)_x \mathrm{d} r$, separate equations for $\tilde{c}$ and $\tilde{u}$ must  be derived.  The equation for $\tilde{c}$ is obtained by subtracting (\ref{eq:cm_averaged_final}) from (\ref{eq:cm_decomposed}) to yield
\begin{equation}
\begin{split}
  \tilde{c}_t + \left(\bar{c}\tilde{u}\right)_x + \left(\tilde{c}\bar{u}\right)_x + \left(\tilde{c}\tilde{u}\right)_x - \frac{2}{a^2}\int_0^a r\left(\tilde{c}\tilde{u}\right)_x \mathrm{d} r \\
 + \frac{1}{r}\frac{\p}{\p r}\left(r \bar{c} \bar{v}\right) + \frac{1}{r}\frac{\p}{\p r}\left(r \bar{c} \tilde{v}\right) + \frac{1}{r}\frac{\p}{\p r}\left(r \tilde{c} \bar{v}\right) + \frac{1}{r}\frac{\p}{\p r}\left(r \tilde{c} \tilde{v}\right) \\
  = D\tilde{c}_{xx} + D\frac{1}{r}(r\tilde{c}_r)_r.
\end{split}
  \label{eq:cm_perturbed}
\end{equation}
To proceed further analytically, additional simplifications to equation (\ref{eq:cm_perturbed}) must be invoked. It is first assumed that $\tilde{c}/\bar{c}\ll1$ as earlier discussed.  Next, a dominant balance argument is employed.  The most important term on the right side is $(1/r) (r\tilde{c}_r)_r$ because $(1/r) (r\tilde{c}_r)_r \sim \textit{O}(1/a^2) \gg \tilde{c}_{xx} \sim \textit{O}(1/L^2)$. This term must balance the three dominant terms on the left side. These terms are the second, six and seventh because all other terms are $\textit{O}(\tilde{c})$, which can be neglected when noting that $(1/r)\p (r\tilde{c}v)/\p r \sim \textit{O}(\epsilon u \tilde{c}/a)$ (the sixth term can also be written as $\bar{c}\bar{v}/r$). This argument holds when assuming that the boundary layer near the membrane is negligible as reasoned elsewhere \citep{aldis1988unstirred,jensen2010self,pedley1983calculation,haaning2013efficiency} and  the term $(\tilde{c}\tilde{u})_x$, which, even though averaged, remains smaller than $(\tilde{u}\bar{c})_x$.  Hence, with these arguments, the dominant balance leads to a simplified and solvable equation for the sought $\tilde{c}$ given by
\begin{equation}
  \left(\bar{c}\tilde{u}\right)_x + \bar{c}\frac{v}{r} + \bar{c}\frac{\p \tilde{v}}{\p r} = D\frac{1}{r}(r\tilde{c}_r)_r.
  \label{eq:c_tilde_ODE}
\end{equation}
For the $\tilde{u}$, $v/r$ and $\p \tilde{v}/\p r$ expressions, the result in equation (\ref{eq:velocities}) can be used when noting that $\bar{u} = -(a^2(8\mu)^{-1})\p \bar{p} / \p x$, $u = \bar{u}\left(2 - (2/a^2)r^2\right)$, $\bar{v} = -(7/15) a \bar{u}_x$ and $v = \bar{u}_x \left(r^3/(2a^2) - r\right)$.  Hence, $\tilde{u}$, $v/r$ and $\p \tilde{v}/\p r$ can now be solved as a function of the area-averaged velocity as $\tilde{u} = u - \bar{u} = \bar{u}\left(1 - (2/a^2)r^2\right)$, $v/r = \bar{u}_x\left(r^2/(2a^2) - 1\right)$, $\p \tilde{v}/\p r = \bar{u}_x\left(r^2(3/2a^2) - 1\right)$ and $\tilde{u}_x = \bar{u}_x\left(1 - (2/a^2)r^2\right)$.  From this result, and noting that the area averaged concentration is only a function of axial position and time, equation (\ref{eq:c_tilde_ODE}) is now separable in radial and axial positions and can be solved for $\tilde{c}$ by integrating in \textit{r} to obtain
\begin{equation}
  \tilde{c} = \frac{1}{D}~\bar{u}~\bar{c}_x~\left(\frac{r^2}{4} - \frac{1}{8 a^2}r^4\right) - \frac{1}{4D}~\bar{u}_x~\bar{c}~r^2 + A(x,t)\ln{r} + B(x,t),
  \label{eq:c_tilde}
\end{equation}
where \textit{A} and \textit{B} are integration constants to be determined.  For  the concentration to be bounded at $r = 0$ it is required that $A = 0$. The area-averaged radial concentration is zero by its definition (i.e. $\int_0^a \left(r\tilde{c}\right) \mathrm{d} r = 0$) leads to $B = (a^2/8D)\bar{u}_x\bar{c} - (a^2/12D)\bar{u}\bar{c}_x$.  Hence, the approximated $\tilde{c}$ and its derivative in the axial position can now be defined as a function of the area averaged concentration and axial velocity using
\begin{equation}
  \tilde{c} = \frac{1}{D}\left(\frac{r^2}{4} - \frac{1}{8a^2}r^4 - \frac{a^2}{12}\right) \bar{u}\bar{c}_x+ \frac{1}{D}\left(-\frac{r^2}{4} + \frac{a^2}{8}\right)\bar{u}_x \bar{c}. 
  \label{eq:c_tilde_result}
\end{equation}
From the fluctuating concentration given in equation (\ref{eq:c_tilde_result}) and the fluctuating velocity given by equation (\ref{eq:velocities}), the integral in equation (\ref{eq:cm_averaged_final}) can now be determined to include the Taylor dispersion effect. After some algebra, the new form of the area averaged equation for conservation of scalar mass can be shown to reduce to
\begin{equation}
  \frac{\p \bar{c}}{\p t} + \frac{\p}{\p x}\left[\left(1 + \frac{a^2}{24 D}\frac{\p \bar{u}}{\p x}\right)\bar{c} \bar{u}\right] = \frac{\p}{\p x}\left[\left(\frac{a^2 \bar{u}^2}{48 D} + D\right)\frac{\p \bar{c}}{\p x}\right].
  \label{eq:cm_TD}
\end{equation}
A number of features in equation (\ref{eq:cm_TD}) should be pointed out when comparing it to the closed-pipe case of Taylor dispersion. The conventional Taylor dispersion term ($=D_d$) is recovered on the right-hand side of equation (\ref{eq:cm_TD}). This $D_d$ is always positive and must act to enhance the apparent diffusion coefficient ($D_d+D$). However, a new term emerges on the left-hand side of equation (\ref{eq:cm_TD}) that is entirely absent in closed-wall pipes. This term acts as an apparent local source or sink of $\bar{c}$ - and its sign is problem and position depended because the mean velocity gradient can be either negative or positive depending on whether the flow is accelerating or decelerating.
\par
The second equation needed to close the problem in the area-averaged form is the membrane physics equation (\ref{eq:membrane_BC}). In this equation, the radial velocity \textit{v} at $r = a$ can be formulated as a function of the area-averaged axial velocity, $v|_{r = a} = - (a/2)\bar{u}_x$. The concentration at the boundary $c_b$ can be simplified by taking it equal to the area-averaged concentration $\bar{c}$. This simplification ignores any boundary-layer effects at the membrane though it abides by pragmatic considerations that $k$ is experimentally determined using averaged quantities when applying a pressure and measuring the average axial velocity. The implication of this assumption is further discussed in appendix \ref{appA}. After differentiating in $x$ to relate the pressure term to the area-averaged axial velocity equation (\ref{eq:membrane_BC}) can be written in the following form:
\begin{equation}
  R_g T \frac{\p \bar{c}}{\p x} = \frac{a}{2 k}\frac{\p^2 \bar{u}}{\p x^2} - \frac{8 \mu}{a^2}\bar{u}.
  \label{eq:membrane_physics}
\end{equation}
Equation (\ref{eq:cm_TD}) can now be used coupled with equation (\ref{eq:membrane_physics}) to offer a new closed-form expression that describes axial sucrose transport in the phloem while accounting for Taylor dispersion.

\section{Simplified Model}
\label{sec:simplified_model}
The findings from equations (\ref{eq:cm_TD}) and (\ref{eq:membrane_physics}) are now interpreted in the context of prior 1-D (axial) theories of phloem transport  \citep{thompson2003application,jensen2009osmotically,huang2018transport}.  As discussed in section \ref{sec:theory}, the tube geometry with $\epsilon\ll1$ can be used to show that the area-averaged concentration $\bar{c}$, the area-averaged pressure $\bar{p}$ and the area-averaged axial velocity $\bar{u}$ are only a function of axial position \textit{x} and time \textit{t}. Prior models commence with the assumption that $\textit{v}/\textit{u}\ll1$ so that the Navier Stokes equation for $v$ becomes un-necessary. Using this assumption, the area-averaged axial velocity is then directly related to the area-averaged pressure gradient (for low $\Rey$ and neglecting gravitational forces) by the HP approximation
\begin{equation}
  \bar{u} = - \frac{a^2}{8 \mu}\frac{\p \bar{p}}{\p x},
  \label{eq:HP}
\end{equation}
as shown in section \ref{sec:point_equations} and discussed elsewhere \citep{thompson2003application, jensen2009osmotically}. The membrane physics can be described using conservation of mass for a constant $\rho$ around a small part of the tube length $\Delta x$, where the osmotic potential and pressure potential are balanced to create an advection difference across $\Delta x$ between inlet position $i$ and outlet position $i+1$ given as \citep{jensen2009osmotically}
\begin{equation}
  \upi a^2 (\bar{u}_{i+1} - \bar{u}_{i}) + 2 \upi a \Delta x k (\Pi - \bar{p}) = 0.
  \label{conservation_volume_1}
\end{equation}
As before, for small $c$, the van't Hoff relation $\Pi = \textit{R}\textit{T}\bar{c}$ can be used to relate the osmotic potential $\Pi$ to $\bar{c}$. Taking the limit $\Delta\textit{x}\ttz$, a relation between $\bar{c}$, $\bar{u}$ and $\bar{p}$ can now be derived and is given by
\begin{equation}
  \frac{a}{2} \frac{\p \bar{u}}{\p x} = k (R_g T \bar{c} - \bar{p}).
  \label{conservation_volume_2}
\end{equation}
This expression arises only from membrane physics and provides a second relation among the three sought variables $\bar{u}$, $\bar{c}$, and $\bar{p}$, which is the same as equation (\ref{eq:membrane_physics}) that was derived from the boundary condition (equation (\ref{eq:membrane_BC})) in section \ref{sec:TD}.  To mathematically close the problem, a third expression between these three variables must be supplied.  However, before presenting this expression, it is to be noted that the HP approximation and membrane physics expressions subject to the van't Hoff approximation are both linear and area-averaged quantities. This linearity breaks down in the scalar mass balance. Starting with the advection-diffusion equation (\ref{eq:advection_diffusion}) and applying the area-averaging operation leads to  
\begin{equation}
  \frac{\p \bar{c}}{\p t} + \frac{\p \overline{u c}}{\p x} = D \frac{\p^2 \bar{c}}{\p x^2}.
  \label{conservation_mass}
\end{equation}
Using equations (\ref{eq:HP}), (\ref{conservation_volume_2}) and (\ref{conservation_mass}), the three equations with three unknowns can be solved (numerically and under some conditions analytically) when setting $\overline{u c}=\overline{u}~ \overline{c}$.  This approximation, which ignores Taylor dispersion, has been used extensively in prior representation of sucrose transport in the phloem \citep{jensen2016sap,jensen2012analytic,thompson2003application,huang2018transport}.  Its consequence is that area-averaged quantities also satisfy the same scalar conservation equation. The inclusion of Taylor dispersion (i.e. arising from $\overline{u c}\neq \overline{u}~ \overline{c}$) in the aforementioned system of equations and tracking its consequences on sucrose front speed frames the compass of the work here.  Equations (\ref{eq:HP}) and (\ref{conservation_volume_2}) can be combined to eliminate $\bar{p}$ resulting in
\begin{equation}
  R_g T \frac{\p \bar{c}}{\p x} = \frac{a}{2 k}\frac{\p^2 \bar{u}}{\p x^2} - \frac{8 \mu}{a^2}\bar{u}.
  \label{membrane_physics}
\end{equation}
The terms on the right-hand side of equation (\ref{membrane_physics}) can be combined to yield a \munch\ number $M$. This is a dimensionless number describing the forces responsible for the axial variation of $\bar{c}$ \citep{jensen2009osmotically}. With ${\p^2 \bar{u}}/{\p x^2}\sim \bar{u}/L^2$, the ratio of the last two terms in equation (\ref{membrane_physics})  result in $M = 16 \mu L^2 k a^{-3}$.

\section{Non-dimensional form for both models}
\label{sec:nondimensional}
This section presents the non-dimensional form for the simplified model derived by \citet{jensen2009osmotically} and discussed in section \ref{sec:simplified_model} and the model that includes Taylor dispersion derived in section \ref{sec:TD}. Because the non-dimensional forms are used to construct model runs for the discussion, they are featured here for convenience depending on whether $M\tti$ or not.

\begin{table}
  \begin{center}
\def~{\hphantom{0}}
  \begin{tabular}{lccccc}
      Runs                                      & 1               &   2             & 3               & 4               & 5\\
      Mean sugar concentration, \^c (mM)        & 1.5 $\pm$ 0.3   & 2.1 $\pm$ 0.03  & 2.4 $\pm$ 0.2   & 4.2 $\pm$ 0.7   & 6.8 $\pm$ 0.1\\
      Osmotic pressure, $\Pi$ (bar)             & 0.14 $\pm$ 0.02 & 0.15 $\pm$ 0.01 & 0.31 $\pm$ 0.03 & 0.39 $\pm$ 0.01 & 0.68 $\pm$ 0.02\\
      Membrane tube length, $L$ (cm)            & 28.5            & 20.8            & 28.5            & 28.5            & 20.6\\
      Initial front height, $l$ (cm)            & 4.9             & 3.7             & 6.6             & 6.5             & 4.8\\
  \end{tabular}
  \caption{Published experimental conditions for the five runs analyzed here.  The reported $R_g T = 0.1$ bars (mM)$^{-1}$.}
  \label{tab:data}
  \end{center}
\end{table}

\subsection{Non-dimensional form of the simplified model}
\label{sec:nondimensional_simplified_model}
Assuming that ${\p \overline{u c}}/{\p x} = {\p \bar{u} \bar{c}}/{\p x}$ (i.e.  radial variation in $u$ and $c$ in the nonlinear term are uncorrelated) and choosing the following scaling for the concentration, velocity, length, and time $c = c_0 C, u = u_0 U, x = L X, t = t_0 \tau$ with
$c_0$ being the initial concentration released at $x=0$, and $u_0$ and $t_0$ are the velocity and time scales to be determined from the problem, equations (\ref{conservation_mass}) and (\ref{membrane_physics}) can be made non-dimensional and given by
\begin{equation}
  \frac{\p C}{\p X} = \frac{\p^2 U}{\p X^2} - M U
  \label{eq:cv_dimensionless_finite_M}
\end{equation}
\begin{equation}
  \frac{\p C}{\p \tau} + \frac{\p U C}{\p X} = \frac{1}{Pe_l} \frac{\p^2 C}{\p X^2},
  \label{eq:cm_dimensionless_finite_M}
\end{equation}
where $t_0 = L u_0^{-1}$ is the advection time scale, $u_0 = 2 k R_g T c_0 L a^{-1}$ is the advection velocity, and $\Pen_l = u_0 L/D$ is the P\'eclet number in the axial direction that can be significant for high $Sc$ even when $u_0$ is small. It is to be noted that this nondimensional number is the inverse of $\bar{D}$ used by \citet{jensen2009osmotically}.  In the limiting case where $M$ is very large, the nondimensional variable $U$ in equation  (\ref{eq:cv_dimensionless_finite_M}) can be re-scaled by $M$ to yield
\begin{equation}
  \frac{\p C}{\p X} =  \frac{1}{M}\frac{\p^2 \hat{U}}{\p X^2} - \hat{U},
  \label{eq:cv_dimensionless_High_M}
\end{equation}
\begin{equation}
    \frac{\p C}{\p \tau} + \frac{1}{M}\frac{\p \hat{U}C}{\p X} = \frac{1}{\Pen_l}\frac{\p^2 C}{\p X^2}
    \label{eq:cm_dimensionless_high_M}
\end{equation}
where $U = \hat{U}/M$ and $\hat{U} = \textit{O}(1)$. When $M\tti$, and using the following expansions for $C$ and $\hat{U}$: $C\sim C_0 + M^{-1} C_1 + \textit{O}(M^{-2}), \hat{U}\sim \hat{U}_0 + M^{-1} \hat{U}_1 + \textit{O}(M^{-2})$, the leading order axial velocity becomes entirely driven by the mean concentration gradient ($\hat{U}_0 = - \p C_0/\p X$), which implies that the conservation of scalar mass becomes independent of the velocity and reduces to a diffusional form characterized by a nonlinear diffusion coefficient given as
\begin{equation}
  \frac{\p C_0}{\p \tau}  =  \frac{\p }{\p X}\bigg[\left (\frac{1}{M} C_0 + \frac{1}{\Pen_l} \right) \frac{\p C_0}{\p X}\bigg].
  \label{eq:cm_dimensionless_high_M_leading_order}
\end{equation}
The leading order analytical solution (when $M^{-1}C \gg 1/\Pen_l$) for this equation can be found elsewhere \citep{jensen2009osmotically} and follows well-established methods for solving such non-linear diffusion problems \citep{king1986diffusion}.

\subsection{Nondimensional Form of the Taylor Dispersion Expression}
\label{sec:TD_nondimensional}
Using the same scaling analysis as before, the equation for the membrane physics is not affected by the derivation of the Taylor dispersion (as expected). However, the nondimensional form for the conservation of scalar mass must be revised to include the radial P\'eclet number $\Pen_r$.  This revision yields
\begin{equation}
  \frac{\p C}{\p \tau} + \frac{\p}{\p X}\left[\left(1 + \frac{\Pen_r}{24}\frac{\p U}{\p X}\right)C U\right] = \frac{\Pen_r}{48} \frac{\p}{\p X}\left[\left(U^2 + \frac{48}{\Pen_r \Pen_l}\right)\frac{\p C}{\p X}\right],
  \label{eq:cm_TD_nondimensional_finite_M}
\end{equation}
where the scaling for the time $t_0$ is the same as in section \ref{sec:nondimensional_simplified_model}. The nondimensional number $\Pen_r = a v_0/D$ defines a radial P\'eclet number where $v_0 = \epsilon u_0$, with $\epsilon = a/L$, as expected from the continuity equation (\ref{eq:continuity}) in section \ref{sec:point_equations}. The nondimensional number $48 \Pen_l^{-1} \Pen_r^{-1}$ is always much smaller than unity leading to $48 \Pen_l^{-1} \Pen_r^{-1} \ll \textit{O}(1)$. However, the nondimensional radial P\'eclet number $\Pen_r/24$ can be large or small depending on the problem and will affect the overall sucrose transport time scale.
\par
As before, in the limiting case where $M$ is very large, the axial velocity can be re-scaled by $M$. In this case, equation (\ref{eq:cm_TD_nondimensional_finite_M}) can be written in the following form:
\begin{equation}
    \frac{\p C}{\p \tau} + \frac{\p}{\p X}\left[\left(\frac{1}{M} + \frac{\Pen_r}{24 M^2}\frac{\p \hat{U}}{\p X}\right)\hat{U} C\right] = \frac{\p}{\p X}\left[\left(\frac{\Pen_r}{48 M^2}\hat{U}^2 + \frac{1}{\Pen_l}\right)\frac{\p C}{\p X}\right]
    \label{eq:cm_TD_nondimensional_high_M}
\end{equation}
For this case, the order of magnitude of the nondimensional number $\Pen_r/(24 M^2)$ will show the importance of the new terms in this model. In the results section, the Taylor dispersion effect for the small to intermediate \munch\ number ($M \ll 1$ or $M =O(1)$) and  large \munch\ number ($M\tti$) cases are presented.

\begin{table}
  \begin{center}
\def~{\hphantom{0}}
  \begin{tabular}{lccccc}
      Runs                                                         & 1      &   2   & 3     & 4     & 5 \\
      Osmotic pressure, $\Pi$ (bar)                                & 0.16   & 0.16  & 0.3  & 0.38  & 0.68 \\
      Permeability, $k \times 10^{-10}$ cm$(Pa s)^{-1}$            & 1.10   & 1.35  & 1.15   & 1.10  & 1.30 \\
  \end{tabular}
  \caption{Different $k$ values needed to match the analytical solution to measurements for each run in figure \ref{fig:comparison_analytical}.}
  \label{tab:data_analytical}
  \end{center}
\end{table}

\section{Results}
The results are divided into two parts. In the first part, a comparison of both the simplified model and the model including Taylor dispersion with published laboratory experiments \citep{jensen2009osmotically} is carried out.  These experiments are in the low \munch\ number regime. From this comparison, indirect evidence of the importance of Taylor dispersion in osmotically driven flows is established. The second part primarily focuses on the role of the new term - the radial P\'eclet number - $\Pen_r$ primarily because $48 \Pen_r^{-1} \Pen_l^{-1} \ll \textit{O}(1)$. That is, molecular diffusion is smaller than the dispersion coefficient for typical phloem dimensions. In each $M$ limit, the behavior of the flow is discussed depending on $\Pen_r$. The work here explores the flow properties and initial conditions affecting the behavior of the sucrose concentration front shape traversing the tube. Flows characterized by small or negligible $M$($\ll 1$) are labeled as HP driven flows whereas flows characterized by very large $M$ are labelled as osmotically driven flow. To be clear, both flow regimes are osmotically driven - and such labeling simply reflects the roles of a fluid property $\mu$ and a membrane property $k$ on the relative magnitudes of the two terms in equation (\ref{membrane_physics}). Further details about the consequences of large or small $M$ on the scaling of $\bar{p}$ is featured in the appendix \ref{appB} for completeness.

\begin{figure}
  \centering
  \includegraphics[width = 0.7\textwidth]{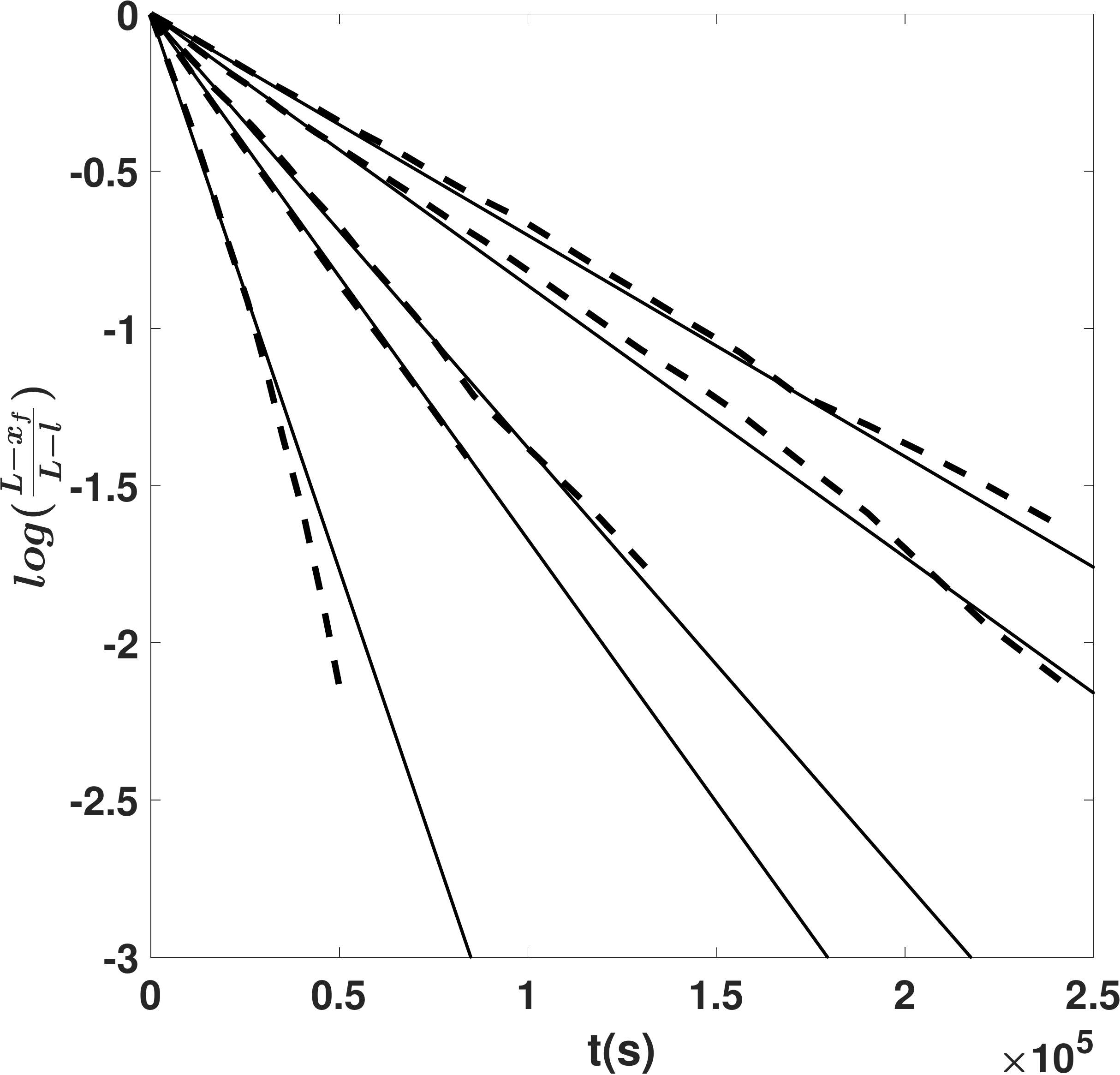}
  \caption{Log plot for the relative front position as a function of time $t$ for the measured concentration (dashed line) and the analytical result given by \citet{jensen2009osmotically} for the low \textit{M} limit (solid line) for the five different experimental runs.}
\label{fig:comparison_analytical}
\end{figure}

\subsection{Comparisons with Published Experiments}
Indirect evidence for the significance of Taylor dispersion in osmotically driven laminar flow is presented based on published experiments. The data used here were extracted from an experiment described elsewhere \citep{jensen2009osmotically} where $M \sim 10^{-8}$. In this experiment, the authors compared an analytical solution derived for very small \textit{M} and $\bar{D} = 1/\Pen_l$ with measurements without including Taylor dispersion in their model.

\subsubsection{Experimental Setup}
\label{experimental_setup}
\begin{figure}
  \centering
  \includegraphics[width = \textwidth]{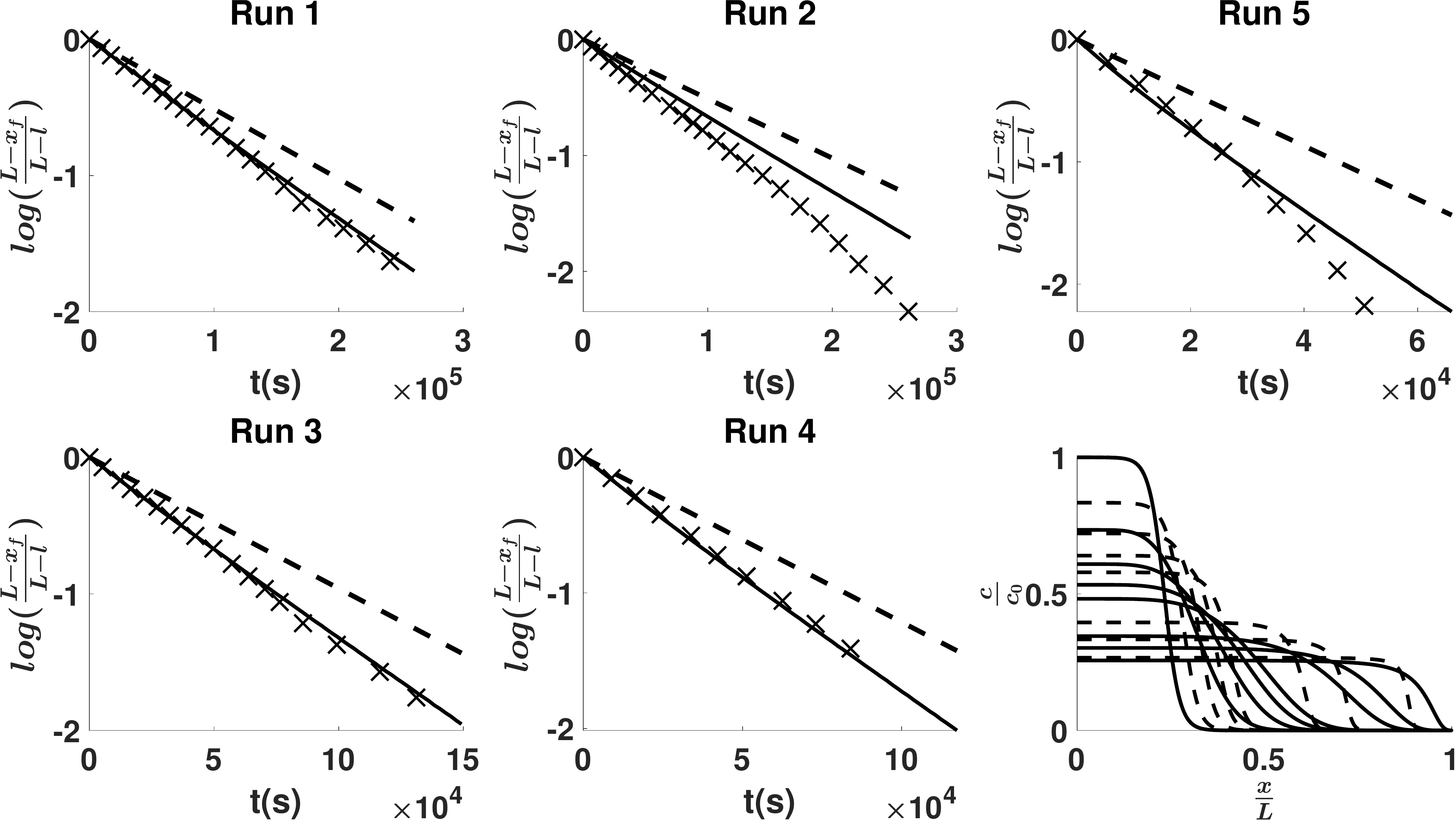}
  \caption{Log plot for the relative front position as a function of time for the experiments (cross), analytical model for the low \textit{M} limit (dashed line) and Taylor dispersion (solid line). The last subplot shows the concentration profile for the fifth run as a function of the axial position for the simplified model developed by \citet{jensen2009osmotically} (dashed line) and the Taylor dispersion model proposed here (solid line).  The inclusion of Taylor dispersion increases the front speed.}
\label{fig:comparison_all_runs}
\end{figure}

The experiment consisted of a tube with $L = 30$ cm and radius $a = 0.5$ cm, with semi-permeable membrane walls to allow  water but not sucrose to be exchanged with the tube. This tube was placed vertically in a water reservoir where the gravitational forces can be assumed to be negligible compared to the pressure gradient. This setup, shown in figure \ref{fig:problem_experiment}, was used for five experimental runs where osmotic pressure and $L$ were varied. The reported constant values for the dynamic viscosity and molecular diffusion in these experiments were $\mu = 1.5 \times 10^{-3}$ Pa s and $D = 6.9 \times 10^{-11}$ m$^2$ s$^{-1}$. Table \ref{tab:data} summarizes the different parameters for the five runs. In all five runs, the two non-dimensional numbers $M$ and $\bar{D}$ ($ = 1/\Pen_l$) are small ($M \sim 10^{-8}$ and $\bar{D} \sim 10^{-5}$) and are neglected in equations (\ref{eq:cv_dimensionless_finite_M}) and (\ref{eq:cm_dimensionless_finite_M}). In the absence of Taylor dispersion, this approximation allowed an analytical result to be derived for the mean concentration front position $x_f(t)$ given by \citep{jensen2009osmotically}
\begin{equation}
  \frac{x_f}{L} = 1 - \left(1-\frac{l}{L}\right)\exp\left(-\frac{t}{t_0}\right),
  \label{eq:front_position_low_M}
\end{equation}
where $l$ is the initial front height at $t=0$ and $t_0 = a (2 k R_g T \hat{c})^{-1} = a (2 k \Pi)^{-1}$.
\par
Figure \ref{fig:comparison_analytical} shows the relative front position $(L - x_f)/(L - l)$ as a function of $t$ for the five runs. To fit their analytical result from (\ref{eq:front_position_low_M}) to experiments, different values for membrane permeability were used.  It is to be noted that $a$ was constant and the osmotic potential was estimated and reported based on concentration measurements.  The membrane material was not altered from run to run, which implies that $k$ ought to be the same across the five runs. The need to vary $k$ across runs led to the conjecture that the term $\overline{(cu)}_x$ may not be $(\bar{c}\bar{u})_x$ and Taylor dispersion may have some impact on this experiment. The different values used to plot figure \ref{fig:comparison_analytical} are shown in table \ref{tab:data_analytical}. Other combinations can be formulated by changing the osmotic potential for each run while changing the permeability. However, no other combination led to a constant permeability for all the five runs. For this reason, we use these values for the model in the following section.

\subsubsection{Data-Model Comparisons}
From section \ref{experimental_setup}, different values of $k$ were necessary to fit the published analytical solution to the measurements for each run. In this section, the model for $x_f(t)$ that includes Taylor dispersion is now used to fit the data but using a single $k$ value across runs. For both models, the permeability $k$ was set to a constant $k = 0.8 \times 10^{-10}$ cm (Pa s)$^{-1}$, which yielded the best fit for all runs (figure \ref{fig:comparison_all_runs}).

\begin{figure}
  \centering
  \includegraphics[width = 0.7\textwidth]{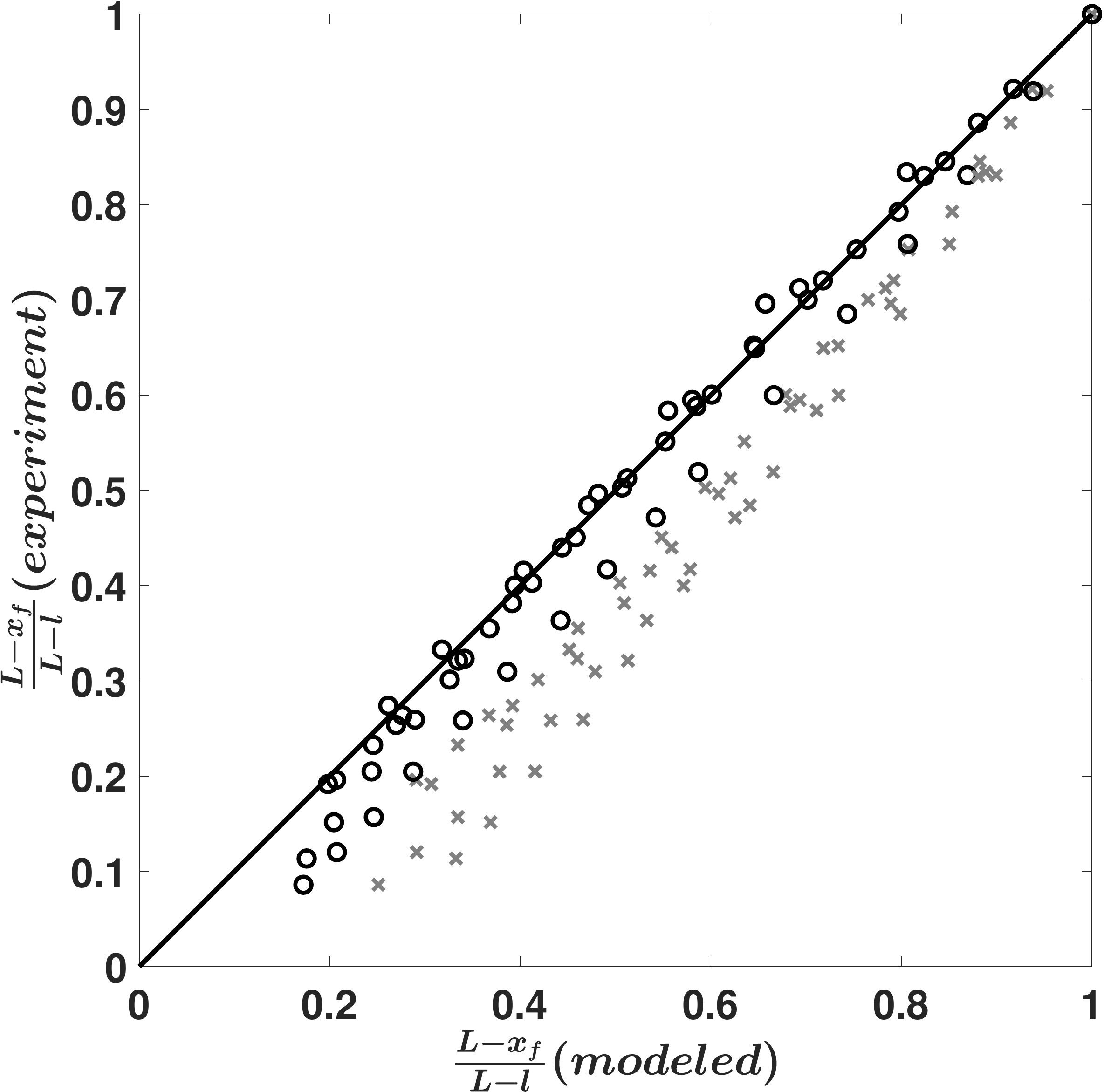}
  \caption{Published data for the relative front position extracted by us from \citet{jensen2009osmotically} as a function of the relative front position calculated from the two different models for the five runs. Circle black points represent the Taylor dispersion model (derived here) and grey crossed points represent the analytical approximation as derived by \citet{jensen2009osmotically}.}
\label{fig:1x1}
\end{figure}

\par
The Taylor dispersion model agrees with measurements for four out of the five runs (figure \ref{fig:comparison_all_runs}).  Only the second run did not agree well with the proposed model for this $k$ value at early times. One possible explanation for this discrepancy is that the measured osmotic potential may have been reported incorrectly since it is related to the mean concentration $\hat{c}$ by $\Pi = R_g T \hat{c}$ (published $R_g T = 0.1$ bars (mM)$^{-1}$ for all runs). From table \ref{tab:data}, when the lower limit for $\hat{c} = 2.07$ mM and the upper limit for the osmotic pressure $\Pi = 0.16$ bar are used, the van't Hoff relation appears not to be satisfied with published $R_g T = 0.1$ bar (mM)$^{-1}$. The osmotic pressure should have attained a higher value, which can increase the osmotic driving force leading to a faster flow and rapid $x_f$ advancement at early times. When revising the driving osmotic pressure to be compatible with the concentration, the agreement with the proposed model here is satisfactory (not shown).
\par
The last plot of figure \ref{fig:comparison_all_runs} presents the axial mean concentration for the fifth run at different time steps. The addition of Taylor dispersion lead to a different shape (not characterized by a front-like behavior) at small time scales. As shown in figure \ref{fig:comparison_all_runs}, the longitudinal mean concentration variation of the Taylor dispersion model is different from the typical wave equation expected at the low $M$ limit. At earlier times, the concentration does not advance with a sharp front because of the highly diffusive behavior at early times. However, at later times, the concentration recovers the expected wave-like front.
\par
Figure \ref{fig:1x1} present a comparison between the relative front position extracted from the data and the relative front position calculated from the two different models (circle points for the Taylor dispersion model and crossed points for Jensen's model) and for the five different runs. The Taylor dispersion model here (with constant $k$) appears to provide a better fit than the one derived without Taylor dispersion. As expected, the relative front position for the second run does not lie on the one-to-one line for reasons linked to the published osmotic potential value earlier described.

\subsection{Results of the models in both regimes}
The effects of Taylor dispersion over a broader range of conditions than those covered by the experiments are now discussed. This discussion is centered on a comparison between the formulation that maintains Taylor dispersion and the standard approach that ignores it. In appendix \ref{appC}, a comparison between these two models and a model that ignores the advection term while maintaining the original Taylor dispersion term $D_d$ as local effect will be presented as well to describe the effect of the advection term on the flow.

\begin{table}
  \begin{center}
\def~{\hphantom{0}}
\begingroup
\setlength{\tabcolsep}{10pt}
\renewcommand{\arraystretch}{1}
  \begin{tabular}{lcccc}
                             & Case 1 & Case 2 & Case 3 & Case 4 \\
      $L$ (m)                & $0.1$                     & $2 \times 10^{-1}$      & $1$                   & $10$ \\
      $a$ (m)                & $10^{-4}$                 & $10^{-3}$               & $1 \times 10^{-5}$    & $1 \times 10^{-4}$ \\
      $k$ m (Pa s)$^{-1}$    & $10^{-12}$                & $10^{-11}$              & $1 \times 10^{-11}$   & $5 \times 10^{-11}$ \\
      $c_0$ mMol             & $10$                      & $20$                    & $10$                  & $40$ \\
      $u_0$                  & $2 \times 10^{-4}$        & $8 \times 10^{-4}$      & $2 \times 10^{-1}$    & $4$ \\
      $t_0$                  & $5 \times 10^2$           & $2.5 \times 10^{2}$     & $5$                   & $2.5$ \\
      \textit{M}             & $2.4 \times 10^{-4}$      & $9.6 \times 10^{-6}$    & $2.4 \times 10^2$     & $1.2 \times 10^2$ \\
      $\epsilon$\Rey         & $1.33 \times 10^{-5}$     & $2.7 \times 10^{-3}$    & $1.3 \times 10^{-5}$  & $2.67\times 10^{-3}$ \\
      $Pe_l$                 & $2.9 \times 10^{5}$       & $2.3 \times 10^{6}$     & $2.9 \times 10^{9}$   & $5.8 \times 10^{11}$ \\
      $Pe_r$                 & $2.9 \times 10^{-1}$      & $5.8 \times 10^{1}$     & $2.9 \times 10^{-1}$  & $5.8 \times 10^{1}$ \\
  \end{tabular}
  \endgroup
  \caption{List of initial conditions and nondimensional numbers for the four different cases discussed. All timescales are advection timescales (including the cases for high $\Pen_r$)}
  \label{tab:cases}
  \end{center}
\end{table}

\par
Now, when designing a broad range of flow conditions (for the finite $\Pen_r$ representation), it is imperative to assess how high $\Pen_r$ can be reached without violating the simplifications to the Navier-Stokes equations (\ref{eq:NSE}). To do so, it is assumed that the highest order of magnitude that the reduced $\Rey$ (i.e. $\epsilon~\Rey$) can reach is $\textit{O}(10^{-2})$. The nondimensional numbers $\epsilon~\Rey$ and $\Pen_r$ can be written as $\epsilon~\Rey = \epsilon \rho u_0 a/\mu = \rho v_0 a/\mu$ and $\Pen_r = v_0 a/D$. This leads to $v_0 a = (\mu/\rho) \epsilon~\Rey = (\mu/\rho) \textit{O}(10^{-2})$, which means that the highest $\Pen_r$ is $v_0 a/D = \mu (\rho D)^{-1} \textit{O}(10^{-2})$. Inserting the values for $\rho$, $\mu$ and $D$, the highest order of magnitude for $\Pen_r$ that can be sustained without the addition of inertial terms in the Navier-Stokes equations is $\textit{O}(10)$. This result implies that the radial advection can be equal to or higher that the radial diffusion.  Obviously, with such high radial advection, the osmotic efficiency might be overestimated \citep{aldis1988unstirred}. The implication of this assumption is further discussed in appendix \ref{appA}.

\subsubsection{Results for HP driven flows}
\label{sec:TD_nondimensional_viscous}
For this type of flow, the pressure gradient is the main driving force and is scaled by viscous forces, hence the name HP driven flow. It either dominates or has similar importance as the osmotic potential. As discussed in section \ref{sec:nondimensional_simplified_model}, \textit{M} is finite (or $M\ttz$) and the velocity is scaled by the boundary condition (i.e. membrane physics), which as shown in appendix \ref{appB}, results in $u_0 = 2 k R_g T c_0 L a^{-1}$. For this case, the two dimensionless numbers in the conservation of scalar mass equation (\ref{eq:cm_TD_nondimensional_finite_M}) can be written in the following forms: $\Pen_r = 2 k R_g T c_0 a D^{-1}$ and $\Pen_l = 2 k R_g T c_0 L^2 a^{-1} D^{-1}$.
\par
In this section, the effect of $\Pen_r$ for the HP driven regime will be discussed. The case where $\Pen_r$ is very small forms a logical starting point for discussion. Its effect on the flow when it reaches the aforementioned upper limit is then analyzed. To do so, a numerical solution using finite difference is obtained for both models. For the Taylor dispersion model, the system of equations solved is equation (\ref{eq:cm_TD_nondimensional_finite_M}) and equation (\ref{eq:cv_dimensionless_finite_M}). For the simplified model, it is equation (\ref{eq:cm_dimensionless_finite_M}) and equation (\ref{eq:cv_dimensionless_finite_M}).
\par

\begin{figure}
    \begin{subfigure}[t]{0.5\textwidth}
        \centering
        \includegraphics[width = \textwidth]{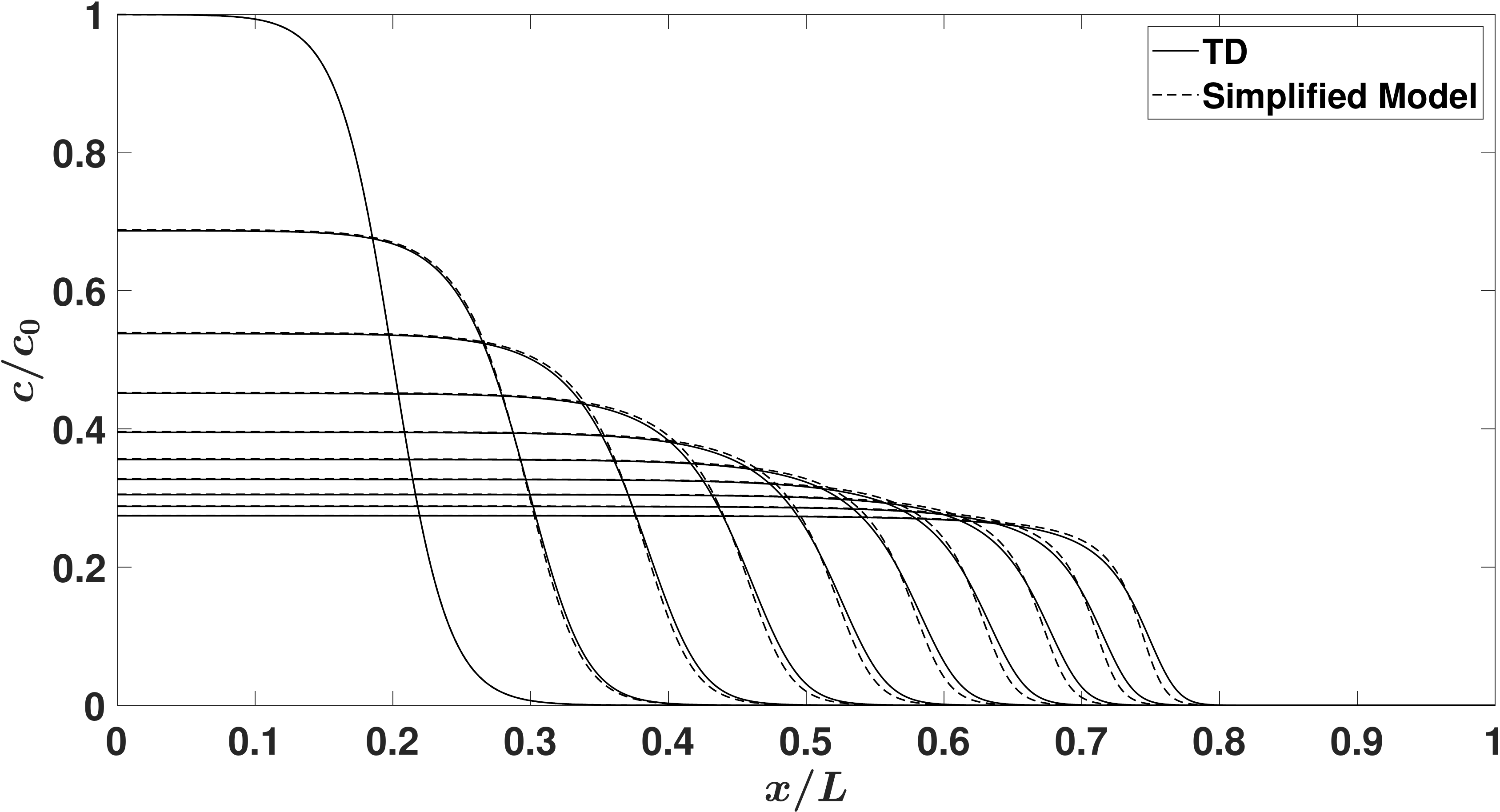}
        \caption{(\textit{a})}
        \label{fig:C_low_M_low_Pe}
    \end{subfigure}
    \begin{subfigure}[t]{0.5\textwidth}
        \centering
        \includegraphics[width = \textwidth]{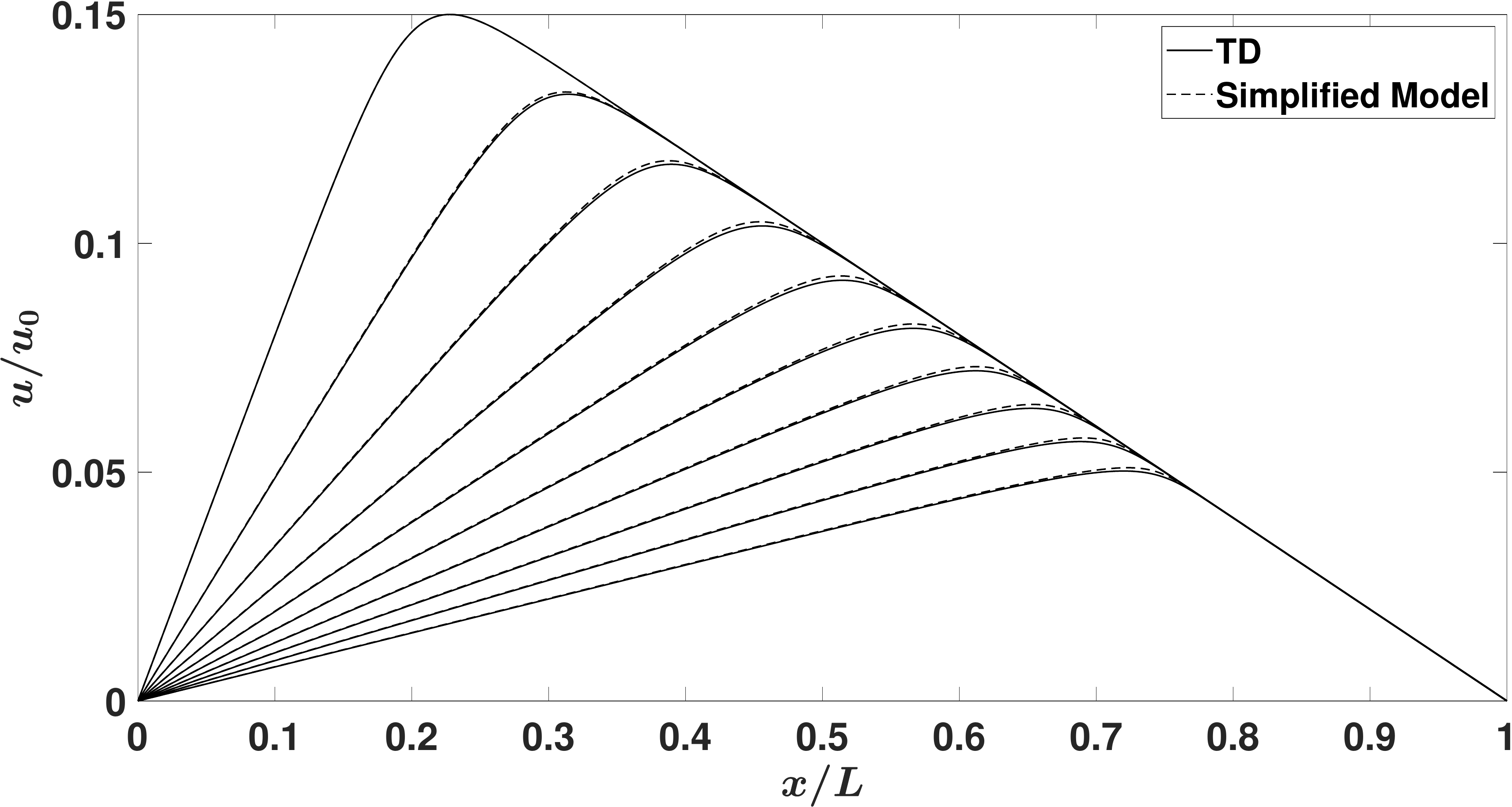}
        \caption{(\textit{b})}
        \label{fig:V_low_M_low_Pe}
    \end{subfigure}
    \begin{subfigure}[t]{0.5\textwidth}
        \centering
        \includegraphics[width = \textwidth]{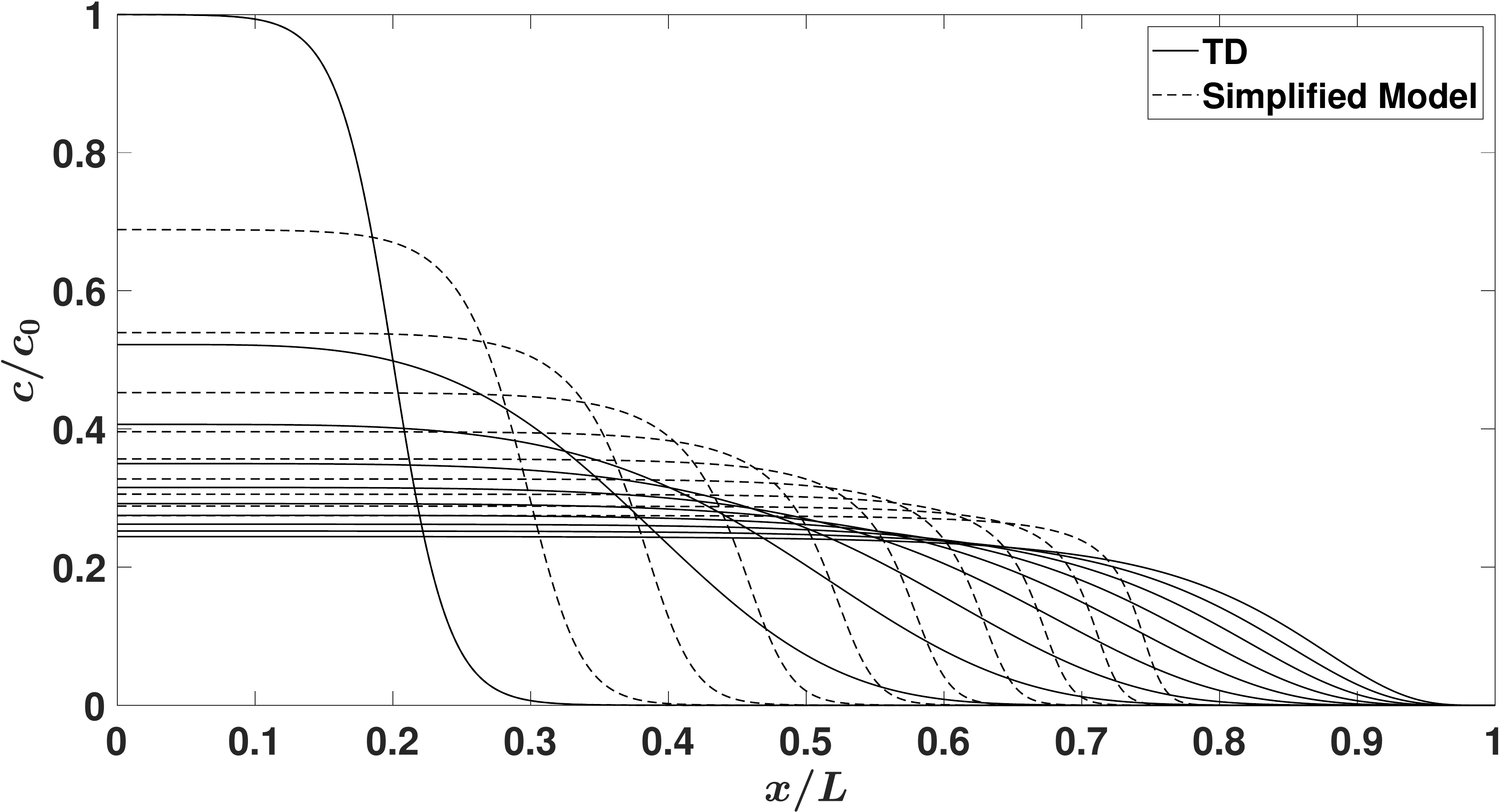}
        \caption{(\textit{c})}
        \label{fig:C_low_M_high_Pe}
    \end{subfigure}
    \begin{subfigure}[t]{0.5\textwidth}
        \centering
        \includegraphics[width = \textwidth]{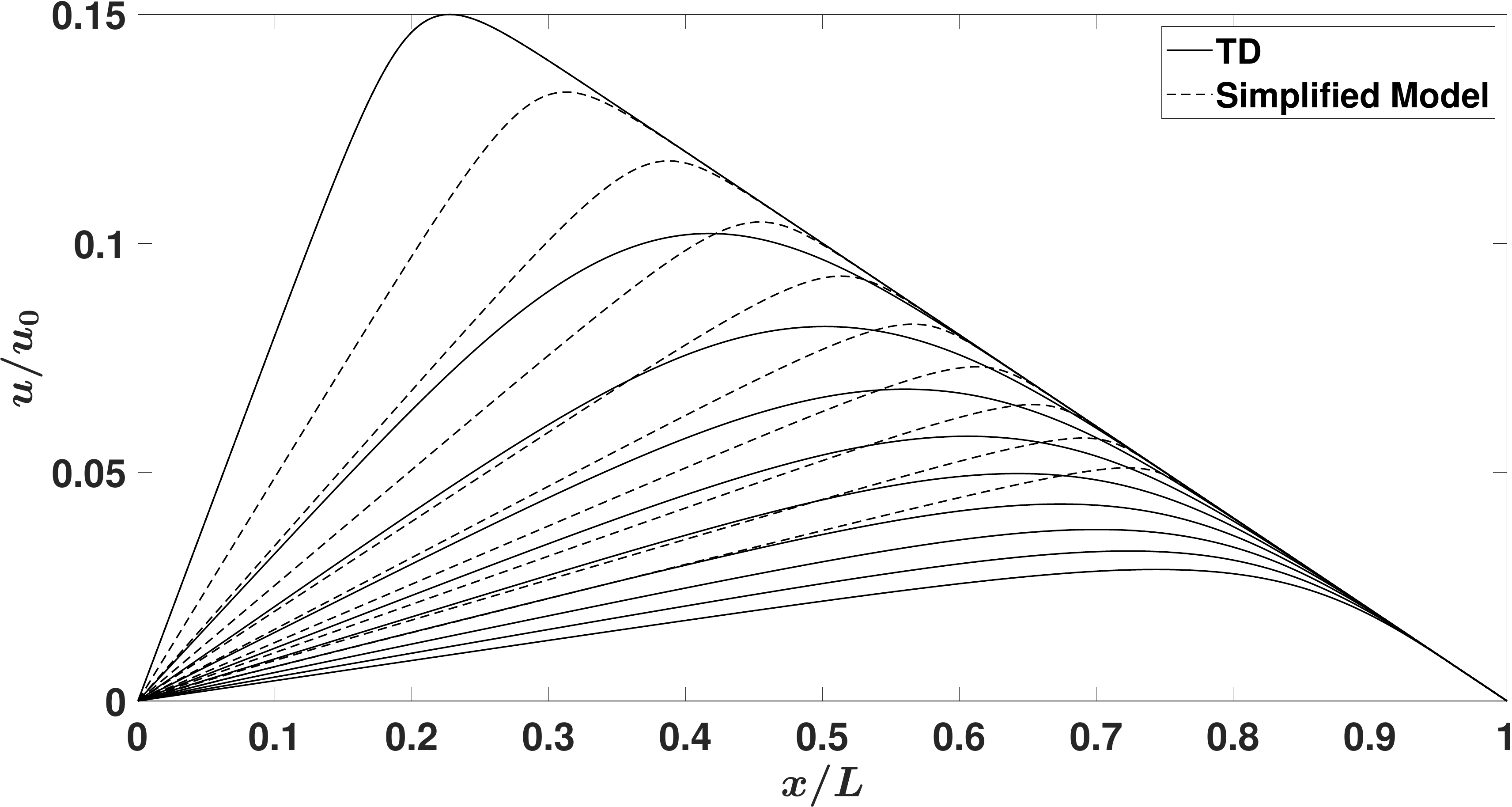}
        \caption{(\textit{d})}
        \label{fig:V_low_M_high_Pe}
    \end{subfigure}
    \caption{Results of the numerical solution for low $M$ number (first and second cases in table \ref{tab:cases}). (\textit{a}) time evolution of the concentration and (\textit{b}) time evolution of the longitudinal mean velocity along the tube length for low $\Pen_r$. (\textit{c}) time evolution of the concentration and (\textit{d}) time evolution of the longitudinal mean velocity along the tube length for high $\Pen_r$. Solid lines denote the result with Taylor dispersion and dashed lines denote the result without Taylor dispersion.}
    \label{fig:low_M}
\end{figure}

In the case where $\Pen_r \ll 1$, the non-dimensional form of the scalar mass conservation (equation (\ref{eq:cm_TD_nondimensional_finite_M})), when the molecular diffusion compared to the dispersion due to advection is neglected, can be written as
\begin{equation}
  \frac{\p C}{\p \tau} + \frac{\p}{\p X}\left[\left(1 - \frac{\Pen_r}{24}\frac{\p U}{\p X}\right)C U\right] = \frac{\Pen_r}{48}\frac{\p}{\p X}\left[U^2 \frac{\p C}{\p X}\right].
  \label{eq:cm_TD_nondimensional_viscous_small_Pe}
\end{equation}
From equation (\ref{eq:cm_TD_nondimensional_viscous_small_Pe}), the effect from Taylor dispersion decreases when $\Pen_r \ttz$. At this limit, the equation exhibits similar behavior as the one derived in section \ref{sec:nondimensional_simplified_model}, equation (\ref{eq:cm_dimensionless_finite_M}) but with two adjustments $1/\Pen_l$ and $\Pen_r/48$ now have different orders of magnitude and the diffusion term itself is different. To illustrate the difference, a set of variables and initial conditions have been selected as presented in table \ref{tab:cases}.
\par
From figures \ref{fig:C_low_M_low_Pe} and \ref{fig:V_low_M_low_Pe}, the effect of Taylor dispersion on front speed appears small. The main difference can be seen in figure \ref{fig:C_low_M_low_Pe} where the front shape is smoother with the inclusion of Taylor dispersion. As mentioned before, the reason behind this 'extra' smoothing is the new dispersion term $D_d$ and its order of magnitude is larger than molecular dispersion (i.e. $\Pen_r/48 \sim \textit{O}(10^{-3}) \gg 1/\Pen_l \sim \textit{O}(10^{-6})$).
\par
The second case of interest is high $\Pen_r$. For this type of flow, advection is large enough to introduce a new behavior in the flow. Assuming that the only variables that can be conveniently changed in an experiment are the dimensions of the tube (i.e. radius \textit{a} and length \textit{L}), its property (i.e. the permeability \textit{k}) and the initial condition (i.e. initial concentration $c_0$), the set of variables chosen for illustration are shown in table \ref{tab:cases}.
\par
This flow exhibits a different shape than the previous illustration. In this case, the advection term (from the analysis in section \ref{sec:TD}) is of the same magnitude as the original advection term and the dispersion term is of order $\textit{O}(1)$. From figures \ref{fig:C_low_M_high_Pe} and \ref{fig:V_low_M_high_Pe}, a finite $\Pen_r$ alters the behavior of the flow and the shape of the advancing concentration front disappears. From figure \ref{fig:C_low_M_high_Pe}, it is also evident that when including Taylor dispersion, the concentration front will sense the 'end of the tube' (downstream conditions) before the case where no Taylor dispersion is present. In this case, to illustrate the speed of the flow, the new non-dimensional equation for the conservation of scalar mass can be re-scaled by $\Pen_r/24$ (while neglecting the molecular diffusion term) to yield
\begin{equation}
  \frac{\p C}{\p \tau} + \frac{\p}{\p X}\left[\left(\frac{24}{\Pen_r} - \frac{\p U}{\p X}\right)C U\right] = \frac{1}{2}\frac{\p}{\p X}\left[U^2\frac{\p C}{\p X}\right],
  \label{eq:cm_TD_nondimensional_viscous_large_Pe}
\end{equation}
where the time scale now is defined by $t_0 = 24 L (\Pen_r u_0)^{-1}$. When $\Pen_r/24 > 1$, the magnitude of the time scale is less than the original advection time scale.

\begin{figure}
    \begin{subfigure}[t]{0.5\textwidth}
        \centering
        \includegraphics[width = \textwidth]{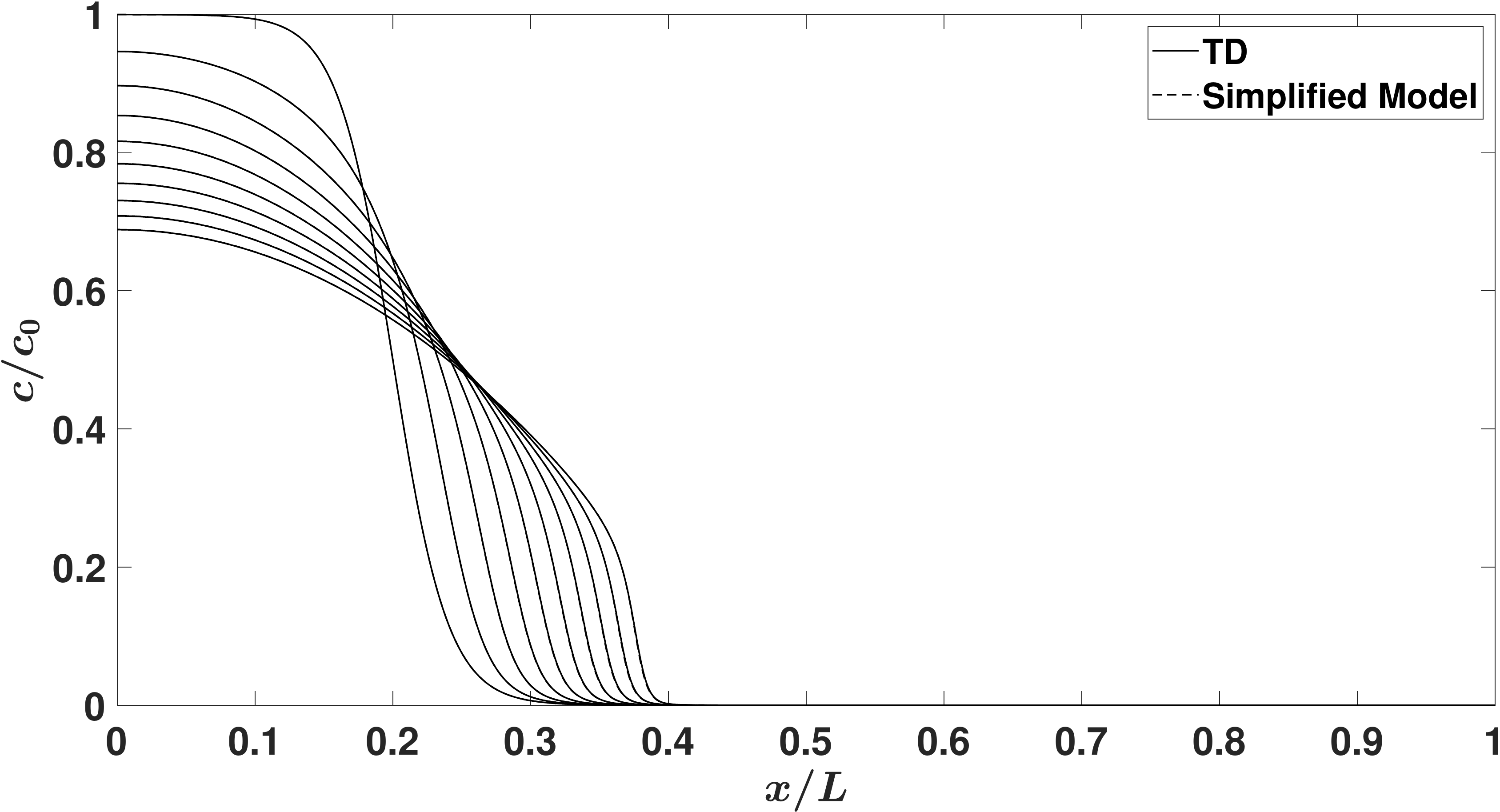}
        \caption{(\textit{a})}
        \label{fig:C_high_M_low_Pe_usingV}
    \end{subfigure}
    \begin{subfigure}[t]{0.5\textwidth}
        \centering
        \includegraphics[width = \textwidth]{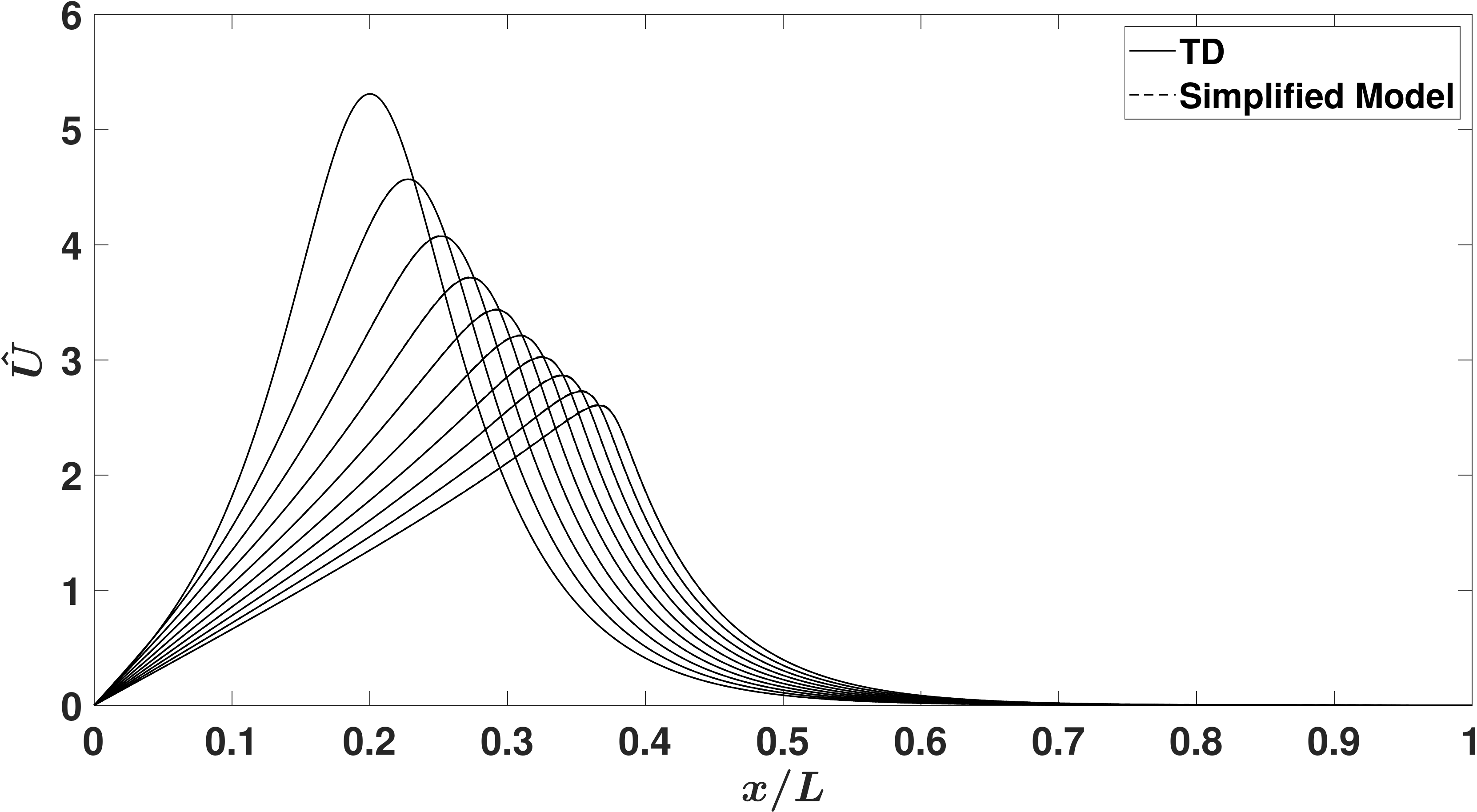}
        \caption{(\textit{b})}
        \label{fig:V_high_M_low_Pe_usingV}
    \end{subfigure}
    \begin{subfigure}[t]{0.5\textwidth}
        \centering
        \includegraphics[width = \textwidth]{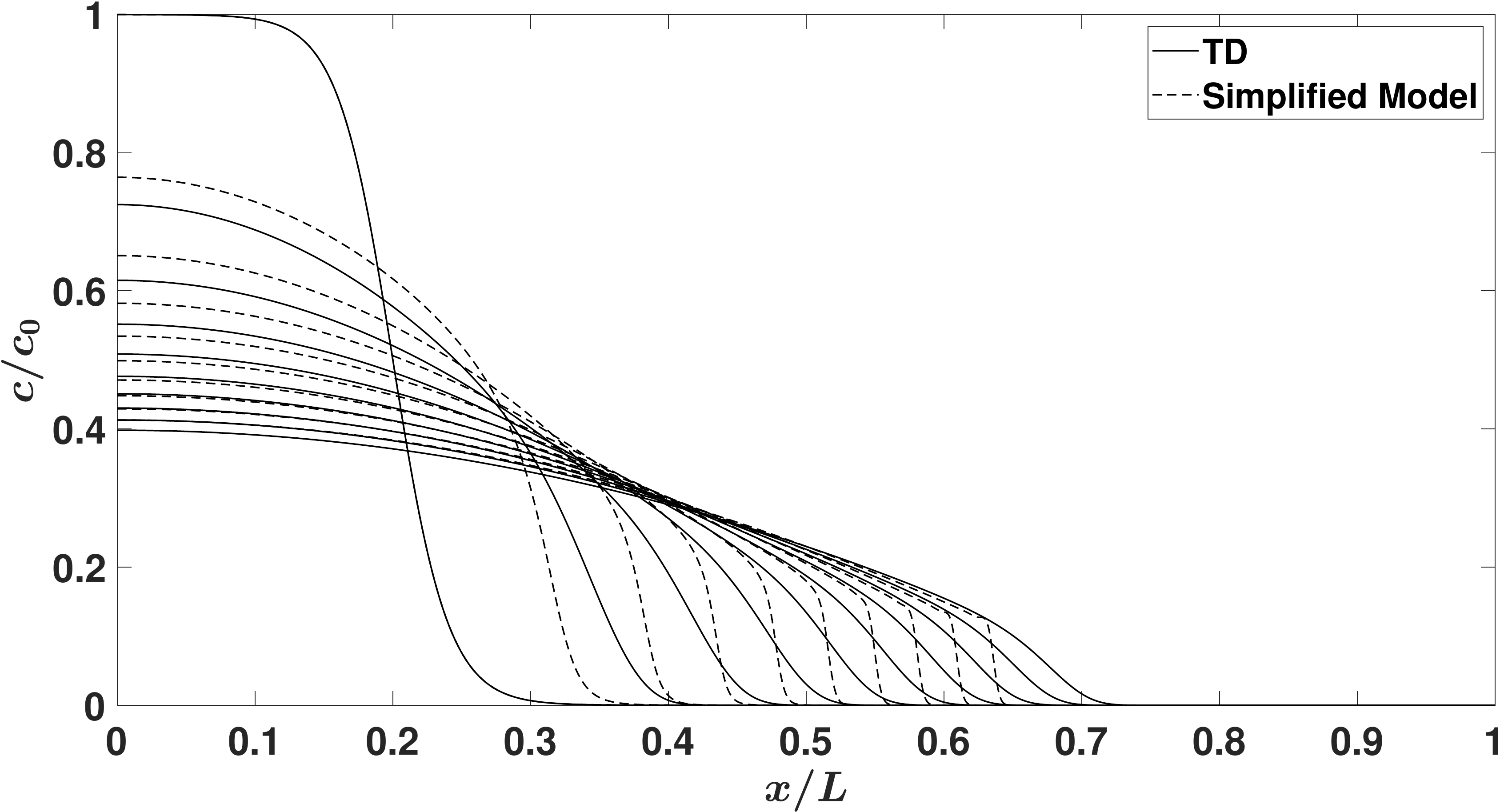}
        \caption{(\textit{c})}
        \label{fig:C_high_M_high_Pe_usingV}
    \end{subfigure}
    \begin{subfigure}[t]{0.5\textwidth}
        \centering
        \includegraphics[width = \textwidth]{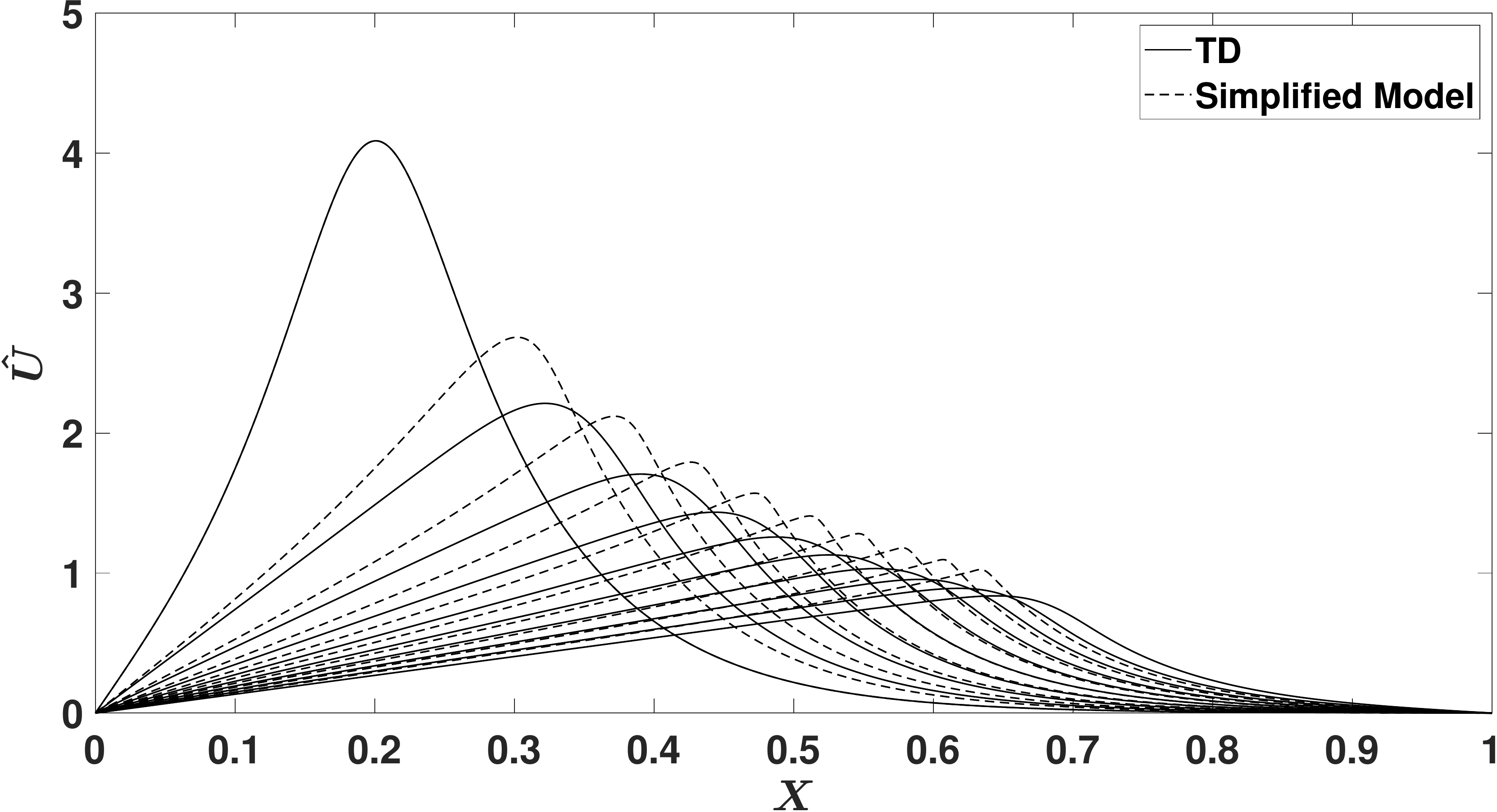}
        \caption{(\textit{d})}
        \label{fig:V_high_M_high_Pe_usingV}
    \end{subfigure}
    \caption{Results of the numerical solution for high $M$ (third and fourth case in table \ref{tab:cases}). (\textit{a}) time evolution of the concentration and (\textit{b}) time evolution of the velocity profile $\hat{U} = M U$ for low $\Pen_r$. (\textit{c}) time evolution of the concentration and (\textit{d}) time evolution of the velocity profile $\hat{U} = M U$ for high $\Pen_r$. Solid lines denote the result with Taylor dispersion and dashed lines denote the result without Taylor dispersion as before.}
    \label{fig:high_M_usingV}
\end{figure}

\subsubsection{Results for osmotically driven flows}
\label{sec:TD_nondimensional_osmotically}
For this type of flow, the pressure can be scaled by the osmotic potential (as shown also in appendix \ref{appB}), hence the name osmotically driven flows. However, to maintain the same scaling for the velocity and pressure as for the case of low $M$, the axial velocity $U$ can be re-scaled by $M$ as discussed in section \ref{sec:nondimensional_simplified_model}. In this case, the system of equations used to obtain the numerical solution is equation (\ref{eq:cm_TD_nondimensional_high_M}) and equation (\ref{eq:cv_dimensionless_High_M}) for the Taylor dispersion model, while equation (\ref{eq:cm_dimensionless_high_M}) and equation (\ref{eq:cv_dimensionless_High_M}) were used for the simplified model. As in section \ref{sec:TD_nondimensional_viscous}, the discussion on the effect of $\Pen_r$ on this type of flow will be presented. The non-dimensional form for the scalar mass is the same as equation (\ref{eq:cm_TD_nondimensional_high_M}). However, to illustrate the importance of Taylor dispersion, the new non-dimensional form can be written using the linear relation between velocity and concentration for very high \textit{M} discussed in \ref{sec:nondimensional_simplified_model}. This non-dimensional form is given as
\begin{equation}
  \frac{\p C_0}{\p \tau} = \frac{\p}{\p X}\left[\left[ \frac{\Pen_r}{48 M^2}\left(\frac{\p C_0}{\p X}\right)^2 + \frac{1}{\Pen_l} + \left(\frac{1}{M} - \frac{\Pen_r}{24 M^2} \frac{\p^2 C_0}{\p X^2}\right) C_0\right]\frac{\p C_0}{\p X}\right],
  \label{eq:cm_TD_nondimensional_osmotically}
\end{equation}
From equation (\ref{eq:cm_TD_nondimensional_osmotically}), one can see that the importance of the new terms resulting from Taylor dispersion can be shown by varying $\Pen_r/M^2$. For this reason, the order of magnitude of $M$ will be approximately the same for both case ( $M \sim \textit{O}(10^2)$), while $\Pen_r$ will be changed from small (i.e $\Pen_r \ll 1$) to its maximum limit (i.e. $\textit{O}(10)$).
\par
For the low $\Pen_r$ limit case, the tube properties and initial conditions, chosen to model this behavior, are shown in table \ref{tab:cases}. As expected, one can see from figures \ref{fig:C_high_M_low_Pe_usingV} and \ref{fig:V_high_M_low_Pe_usingV} that both models behave approximately the same. The reason behind that is the small value for $\Pen_r/M^2$ which leads that both models have the same leading order solution that was discussed in section \ref{sec:nondimensional_simplified_model}. One possible difference can be the analytical solution for the moving boundary layer when $M^{-1}C_0 \ll 1/\Pen_l$, where in this model, the new taylor dispersion terms can have higher order of magnitude in this region. However, this analysis is tangential to the role of Taylor dispersion in the phloem and is better kept for a future inquiry.
\par
As in section \ref{sec:TD_nondimensional_viscous}, the interesting case is the higher $\Pen_r$. For this reason, the set of initial and geometrical conditions, shown in table \ref{tab:cases} with the resulting $\Pen_r$ and $\epsilon\Rey$, are chosen for illustration.
In figures \ref{fig:C_high_M_high_Pe_usingV} and \ref{fig:V_high_M_high_Pe_usingV}, $D_d$ has appreciably smoothed the mean longitudinal velocity and concentration along $x$. Another interpretation for the effect of Taylor dispersion can be seen from the longitudinal concentration distribution in figure \ref{fig:C_high_M_high_Pe_usingV} where the two plots differ in behavior within the vicinity of the scalar moving front. The reason for this difference is evident from equation (\ref{eq:cm_TD_nondimensional_osmotically}) where the analytical solution for the area-averaged concentration along $x$ is dependent on two new terms due to Taylor dispersion (advection and diffusion) unlike the case in equation (\ref{eq:cm_dimensionless_high_M_leading_order}) that depends on a constant term. A simplification can be achieved using an asymptotic analysis for the most of the domain, which ultimately leads to a diffusion coefficient that depends on concentration $C$ only as discussed elsewhere \citep{jensen2009osmotically}. However, for the moving boundary layer, the other terms can be more important. From figures \ref{fig:C_high_M_low_Pe_usingV} and \ref{fig:V_high_M_low_Pe_usingV}, the addition of Taylor dispersion speeds up the self-similar solution compared to the simplified model.
\par
In this type of flow, the axial velocity $U$ scales as $\sim \textit{O}(1/M)$. This means that the radial advection is always much smaller than the radial diffusion even for the case where $\Pen_r > 1$. For this reason, when $M \gg 1$, the effect of Taylor dispersion decreases. However, in appendix \ref{appD}, a scaling analysis will show how the radial P\'eclet number can have a bigger effect if one uses a new scaling for the axial velocity as discussed in appendix \ref{appB} for the case where $M \gg 1$.
\par
The use of the simplified equation (\ref{eq:cm_TD_nondimensional_osmotically}) shows the behavior of the flow in this limit. This equation behaves as a diffusion equation with a diffusion coefficient that depends on the concentration and its derivatives. This apparent diffusion coefficient results from the linear relation between the velocity and the concentration (i.e. $\hat{U}_0 = - \p C_0/\p X$) as a first order approximation. It is the result of molecular diffusion, typical Taylor dispersion and the advection terms. Now, if one looks back to the small $M$ limit, equations (\ref{eq:cm_TD_nondimensional_viscous_small_Pe}) and (\ref{eq:cm_TD_nondimensional_viscous_large_Pe}) behave more like an advection-diffusion equation. In equation (\ref{eq:cm_TD_nondimensional_viscous_small_Pe}), advection dominates the flow and the system resembles a moving wave with a sharp front. In equation (\ref{eq:cm_TD_nondimensional_viscous_large_Pe}), both terms are important, which then leads to a smoother front.

\section{Conclusion}
Description of osmotically driven low Reynolds number flows at high Schmidt numbers within narrow long tubes was revised to include the effects of Taylor dispersion. These flow conditions may arise in the phloem when describing sucrose transport in plants. The conservation of scalar mass suggests that the P\'eclet number, defined by the product of a low Reynolds number and the high Schmidt number, need not be small.  The immediate consequence of such argument is that advective scalar transport is not small necessitating the inclusion of Taylor dispersion. A theory for longitudinal sucrose transport was proposed by area-averaging three inter-related expressions: the Hagen-Poiseuille equation linking velocity and pressure gradients, a membrane physics equation linking velocity gradients to pressure and scalar concentration subject to the van't Hoff approximation, and the advection-diffusion equation for scalar mass linking velocity to concentration.  The dominant balance subject to small deviations in concentration from their area-averaged values allowed explicit governing equations to be derived for the area-averaged pressure, concentration, and velocity.  The Taylor dispersion in the longitudinal direction, originally derived for closed pipes, emerges but with new adjustments due to osmotic effects.  These adjustments act as local sources or sinks of sucrose, though their overall domain-averaged effect is zero.  
The analysis highlighted the unexpected role of a nondimensional radial P\'eclet number $\Pen_r$, which acts upon the area-averaged longitudinal velocity gradient. The work here shows that sucrose transport is faster when adding Taylor dispersion. Hence, for the same sucrose concentration difference between sucrose sources (in leaves) and furthest sinks (in roots), longitudinal sucrose transport in the phloem can be enhanced when including Taylor dispersion. Unlike the original Taylor dispersion in closed pipes that increases the overall apparent longitudinal diffusion, a finite $\Pen_r$ here makes the degree of enhancement problem dependent.  

MN, J-CD, and GK acknowledge support from the U.S. National Science Foundation (NSF-AGS-1644382 and NSF-IOS-1754893). SS was supported by Los Alamos Directed Research and Development Exploratory Research Grant (No. 20160373ER).

\appendix

\section{}\label{appA}
In section \ref{sec:TD}, the area-averaged equation for membrane physics was forced by the concentration at the membrane so that $c_b=\overline{c}(x)$. This assumption is compatible with $\tilde{c}/\overline{c}\ll1$ at $r=a$.  The inclusion of a $\tilde{c}\neq0$ at $r=a$ is tracked here through the membrane physics only and its consequences on the flow is discussed.
\par
The inclusion of $\tilde{c}$ in the membrane physics (linear in $c$) but ignoring its magnitude in the scalar mass balance may be a concern.  To be clear, the objective of this analysis is to illustrate how deviations from $c_b=\overline{c}(x)$ at the boundary impact the final Taylor dispersion theory when $\Pen_r$ is finite. Given that the $c_b$ is likely to be smaller than $\overline {c}$ because the water bath surrounding the tube has a $c=0$, the efficiency of the osmotic potential arising from the membrane physics is likely to be diminished \citep{aldis1988unstirred}. It is this point that is elaborated upon here.
\par

\begin{figure}
    \begin{subfigure}[t]{0.5\textwidth}
        \centering
        \includegraphics[width = \textwidth]{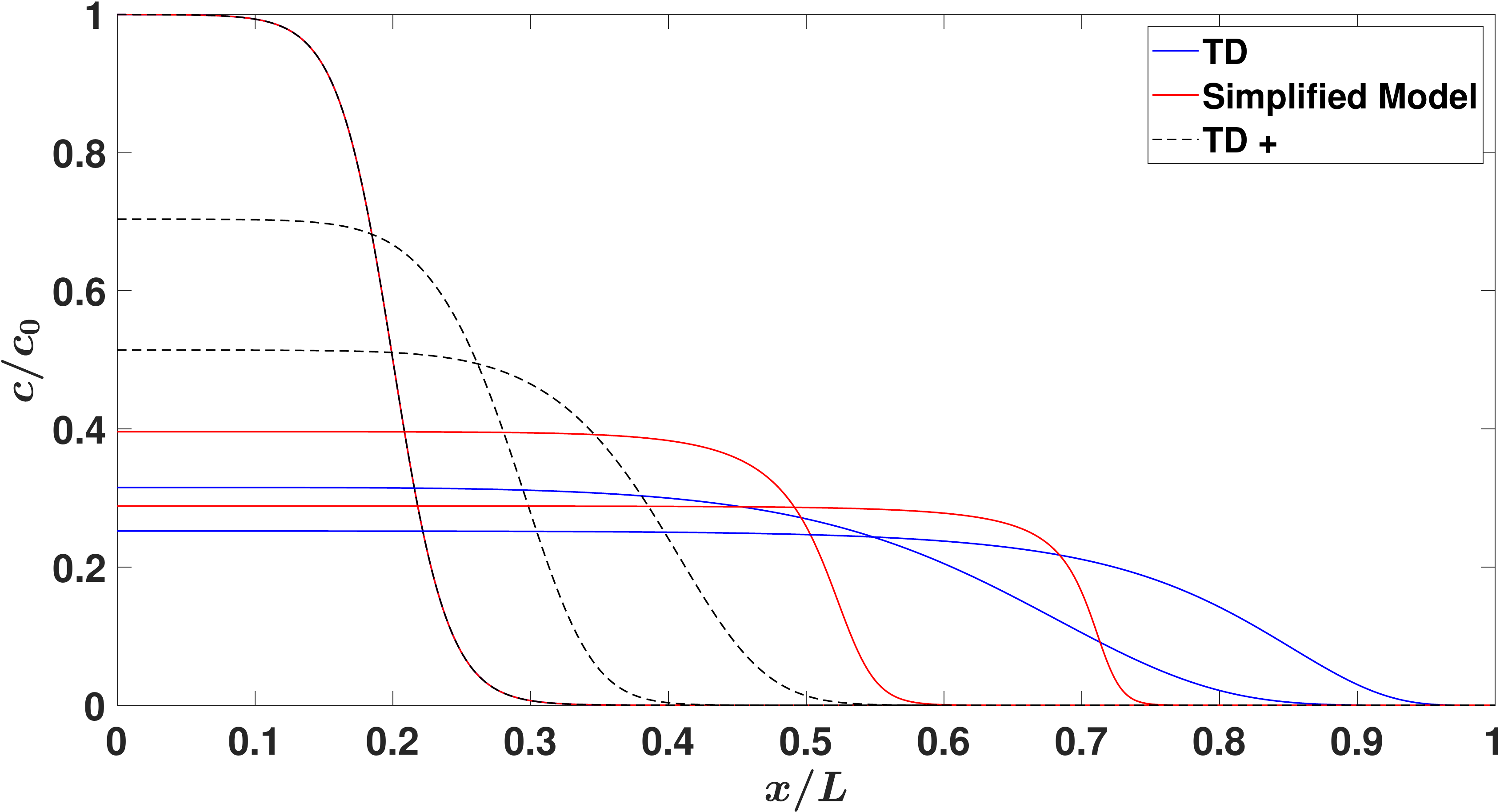}
        \caption{(\textit{a})}
        \label{fig:C_low_M_high_Pe_point}
    \end{subfigure}
    \begin{subfigure}[t]{0.5\textwidth}
        \centering
        \includegraphics[width = \textwidth]{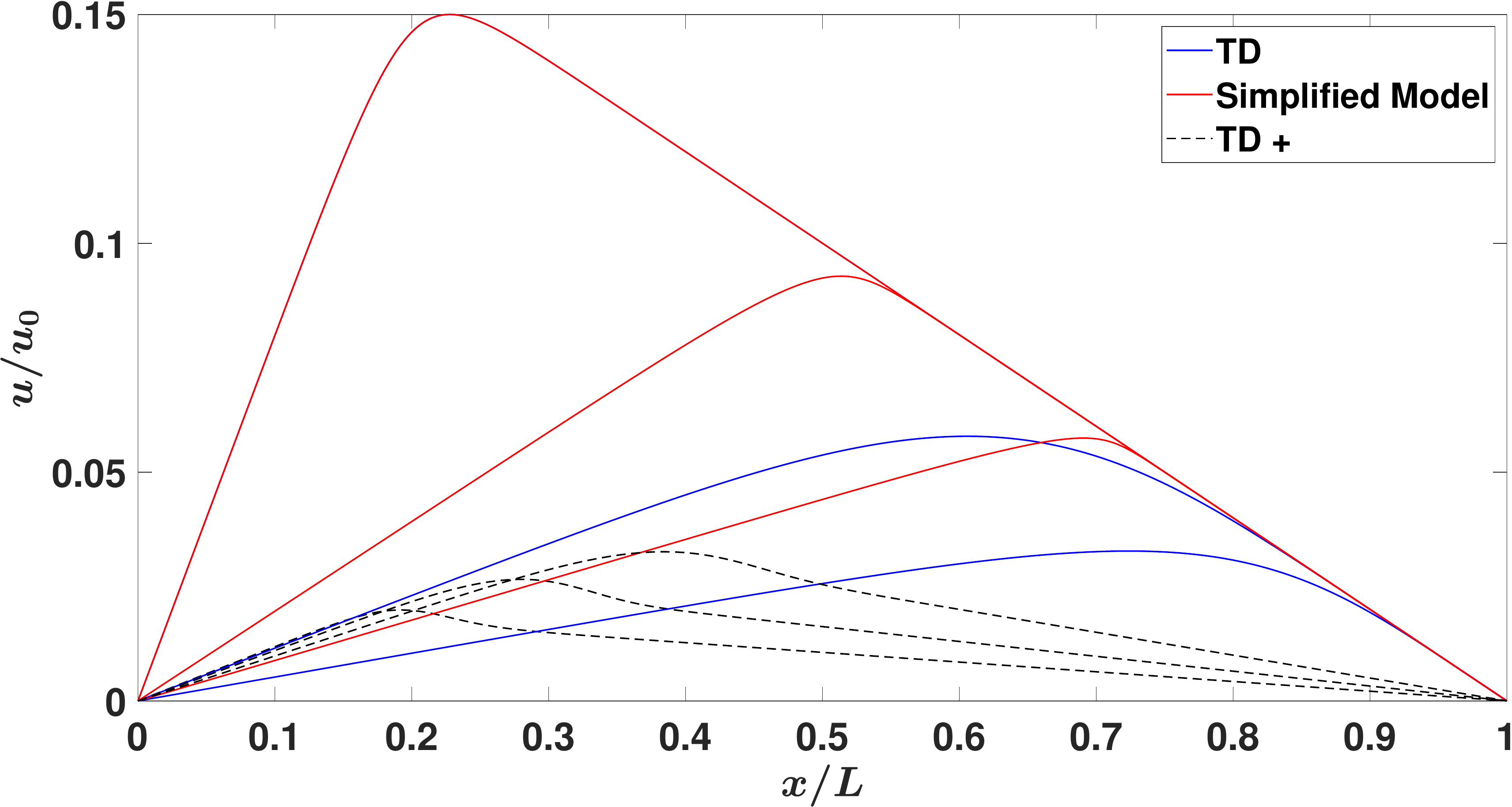}
        \caption{(\textit{b})}
        \label{fig:V_low_M_high_Pe_point}
    \end{subfigure}
    \caption{Results of the numerical solution high $\Pen_r$ and low $M$ for $t = 0$ and at a latter times. (\textit{a}) concentration profile and (\textit{b}) velocity  profile along the tube length for these times. Solid blue lines denote the result with Taylor dispersion (TD), solid red lines denote the result without Taylor dispersion (simplified model) and dashed black lines denote the result with Taylor dispersion including $\tilde{c}$ in the membrane physics labelled as TD$^+$.}
    \label{fig:low_M_high_Pe_point}
\end{figure}

For this reason, the inclusion of $\tilde{c}$ originating from the membrane physics only is tracked - assuming that $\tilde{c}/\overline{c}\ll1$ remains satisfied for much of the remaining tube away from the membrane. In order to show this effect, a comparsion between the Taylor dispersion model derived in section \ref{sec:TD} (denoted by 'TD'), the simplified model derived by \citet{jensen2009osmotically} and summarized in section \ref{sec:simplified_model} (denoted by 'simplified model') and the new model to be derived here, that includes $\tilde{c}$ in membrane physics (denoted by 'TD$^{+}$') will be shown.
\par
Using equation (\ref{eq:c_tilde_result}) for $\tilde{c}$, the membrane physics can be rewritten as
\begin{equation}
    \frac{a}{2 k}\frac{\p^2 \overline{u}}{\p x^2} - \frac{8 \mu}{a^2}\overline{u} = R_g T \left[\frac{\p \overline{c}}{\p x} + \frac{a^2}{24 D}\frac{\p}{\p x}\left(\overline{u}\frac{\p \overline{c}}{\p x}\right) - \frac{a^2}{8 D}\frac{\p}{\p x}\left(\overline{c}\frac{\p \overline{u}}{\p x}\right)\right],
\end{equation}
or in nondimensional form
\begin{equation}
    \frac{\p^2 U}{\p X^2} - M U = \frac{\p C}{\p X} - \frac{\Pen_r}{12}\frac{\p U}{\p X}\frac{\p C}{\p X} + \frac{\Pen_r}{24}U\frac{\p^2 C}{\p X^2} - \frac{\Pen_r}{8} C\frac{\p^2 U}{\p X^2},
    \label{eq:TD_full_nondimensional}
\end{equation}
where, as before, $M$ is the \munch\ number, $\Pen_r$ is the radial P\'eclet number and $u_0$ has the scaling discussed in section \ref{sec:nondimensional_simplified_model} for low $M$. From equation (\ref{eq:TD_full_nondimensional}), the appearance of $\Pen_r$ shows how a high $\Pen_r$ impacts the transport. Physically, when $\Pen_r$ is higher, it is due to a higher radial velocity, which is induced by a higher osmotic potential. In this case, the concentration at the membrane boundary cannot be approximated by the area-averaged equation because the radial advection is higher than the radial diffusion, meaning the deviation from the $\bar{c}$ is not small. This addition slows down the flow because decreasing the concentration at the boundary from  $\bar{c}$ will decrease the osmotic potential.
\par
This conjecture is demonstrated by numerically integrating the full set of equations in $2D$ and then comparing the area-averaged solution with the various approximations invoked.  We choose the case with the highest $\Pen_r$ to illustrate the maximum effects of $c_b<\bar{c}$, and this case is featured in figure \ref{fig:low_M_high_Pe_point}.  For all the low $\Pen_r$ cases, we confirmed that the difference between the two Taylor dispersion approximations and a full 2D numerical solution are minor (not shown).
\par

\begin{figure}
  \centering
  \includegraphics[width = \textwidth]{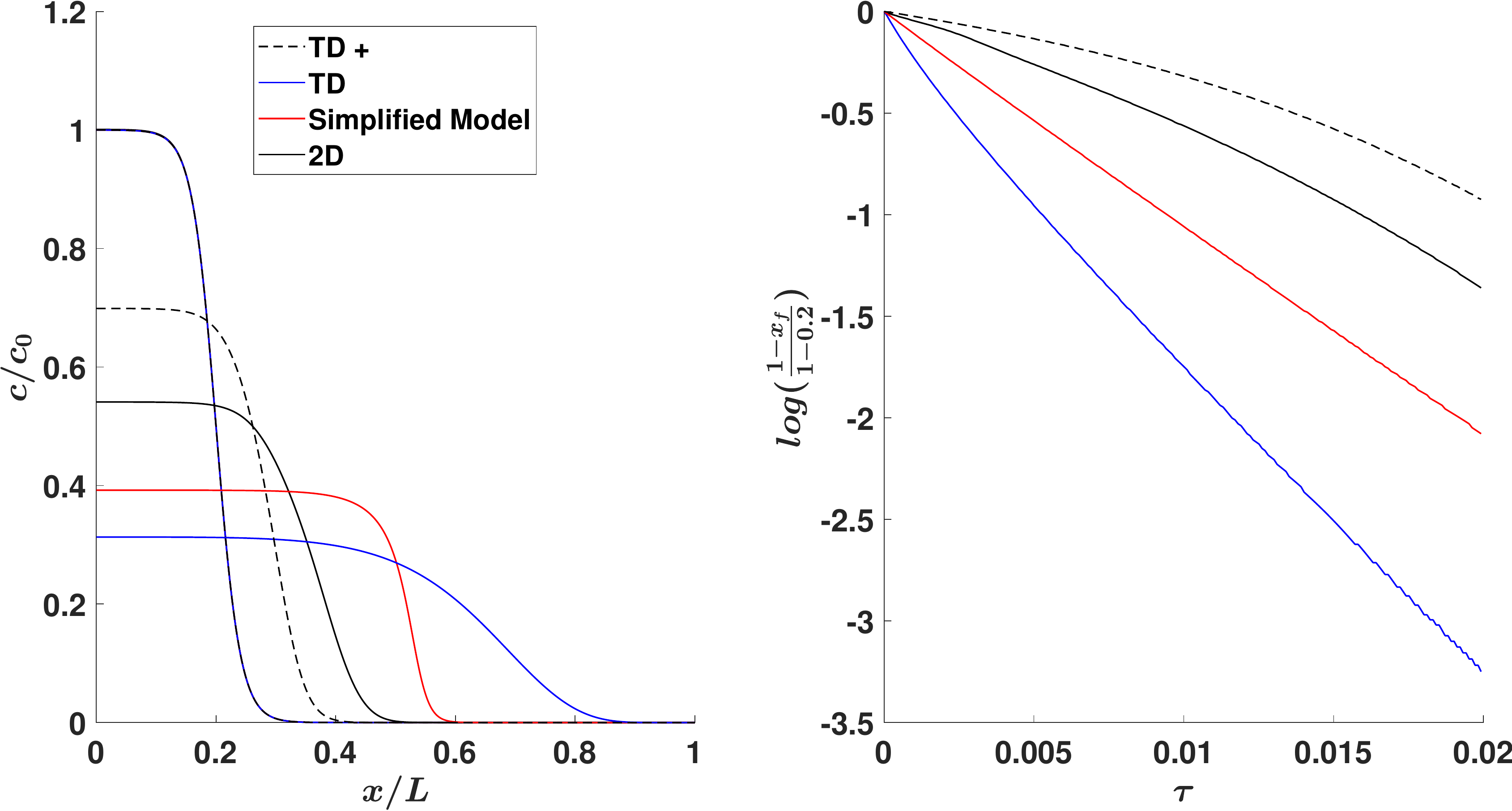}
  \caption{Results of the numerical solution at high $\Pen_r$ and low $M$. (\textit{a}) concentration profile at $t = 0$ and later time and (\textit{b}) time evolution of the logarithmic of the relative front position. Solid black lines denote the results from a numerical integration of the 2D model (i.e. the governing equations), which are then area-averaged and featured here, solid blue lines denote the result with Taylor dispersion, dashed black lines denote the result with Taylor dispersion when including $\tilde{c}$ in the membrane physics (labeled as TD$^+$) and solid red lines denote the result without any Taylor dispersion (simplified model).}
\label{fig:C_front_low_M_high_Pe}
\end{figure}

Figure \ref{fig:low_M_high_Pe_point} illustrates that the use of $\tilde{c}$ derived from equation (\ref{eq:c_tilde_result}) overestimates the slow down of the flow compared to the 2D solution since it was ignored in the scalar mass balance. However, this result is presented to illustrate the tendency of the solution to respond to a reduced concentration at the boundary while preserving the membrane physics. It also implies that the Taylor dispersion model with $c_b=\overline{c}$ is providing an upper limit (compared to the 2D solution) for the sucrose transport speed in such osmotically driven flow as also shown in figure \ref{fig:C_front_low_M_high_Pe}.
\par
From figure \ref{fig:C_front_low_M_high_Pe}\textit{b}, the logarithm of the front position as a function of time is presented. From this figure, one can see that the speed of the 2D model is slower than both model ('TD' and 'simplified') that ignores $\tilde{c}$ in the membrane physics equation. However, the model that includes $\tilde{c}$ overestimates this slow down. To be clear, the TD$^{+}$ and the 2D numerical solution are operating at a reduced osmotic potential driving force when compared to the cases with no TD or the derived TD expression in the main text.  The latter two cases both assume $c_b=\bar{c}$ thereby guaranteeing higher overall velocities in the pipe.
\par
The interesting result in this figure is the deviation from the log-linear shape for the 2D model, which is more apparent in the 'TD$^+$'.  Again, for the lower $\Pen_r$ cases, the various approximations converge and no significant difference can be found.

\section{}\label{appB}
This appendix seeks to clarify the naming of the two flow regimes based on finite (i.e. $M \sim \textit{O}(1)$ or $M \ll 1$) and very large $M$ (i.e. $M\tti$). The two equations needed to obtain the \munch\ number are the membrane physics (i.e. the boundary condition relating the radial flow to the pressure difference across the membrane) and the area-averaged relation between the axial velocity and the pressure (i.e. HP equation). These two equations can be expressed as (after relating $v$ at $r = a$ to $\bar{u}_x$):
\begin{equation}
    \begin{split}
        \frac{a}{2}\frac{\p u}{\p x} = k R_g T c - k p,\\
        u = \frac{a^2}{8 \mu}\frac{\p p}{\p x}.
        \label{eq:naming}
    \end{split}
\end{equation}
Using these equations, it is now shown why the scaling for the axial velocity and pressure is different from one $M$ regime to the other. In the low $M$ limit, the nondimensional form of equations (\ref{eq:naming}) can be written as
\begin{equation}
    \begin{split}
        \frac{\p U}{\p X} = \frac{2 k R_g T c_0 L}{a u_0}C - \frac{2 k L p_0}{a u_0} P,\\
        U = \frac{a^2 p_0}{8 \mu u_0 L}\frac{\p P}{\p X}.
        \label{eq:naming_nondimensional}
    \end{split}
\end{equation}
In equation (\ref{eq:naming_nondimensional}), if the velocity scale is obtained from the osmotic pressure and the pressure from the HP equation, the nondimensional form becomes
\begin{equation}
    \begin{split}
        \frac{\p U}{\p X} = C - M P,\\
        U = \frac{\p P}{\p X},
        \label{eq:naming_nondimensional_low_M}
    \end{split}
\end{equation}
where $u_0 = 2 k R_g T c_0 L a^{-1}$, $M = 16 k \mu L^2 a^{-3}$ and $p_0 = 8 \mu u_0 a^{-3} L = M R_g T c_0$. From this scaling, $M$ is the ratio of the viscous pressure potential to the osmotic potential. The scaling for the pressure here originates from the viscous forces. This case represents a flow that depends on viscosity because the pressure was scaled from the Navier-Stokes equations and leads to a velocity scaling from the boundary condition.
\par

\begin{figure}
    \begin{subfigure}[t]{0.5\textwidth}
        \centering
        \includegraphics[width = \textwidth]{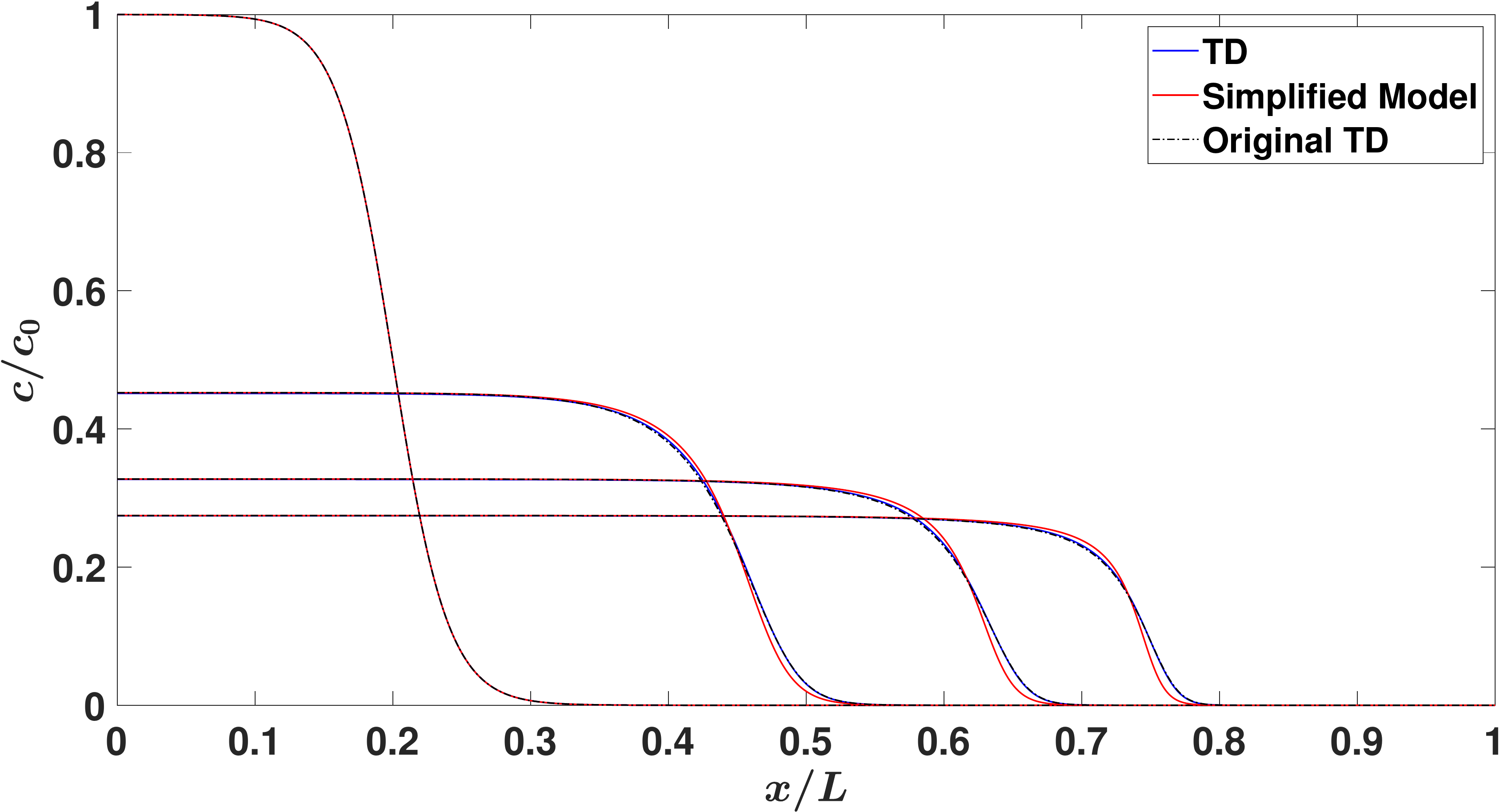}
        \caption{(\textit{a})}
        \label{fig:C_low_M_low_Pe_original}
    \end{subfigure}
    \begin{subfigure}[t]{0.5\textwidth}
        \centering
        \includegraphics[width = \textwidth]{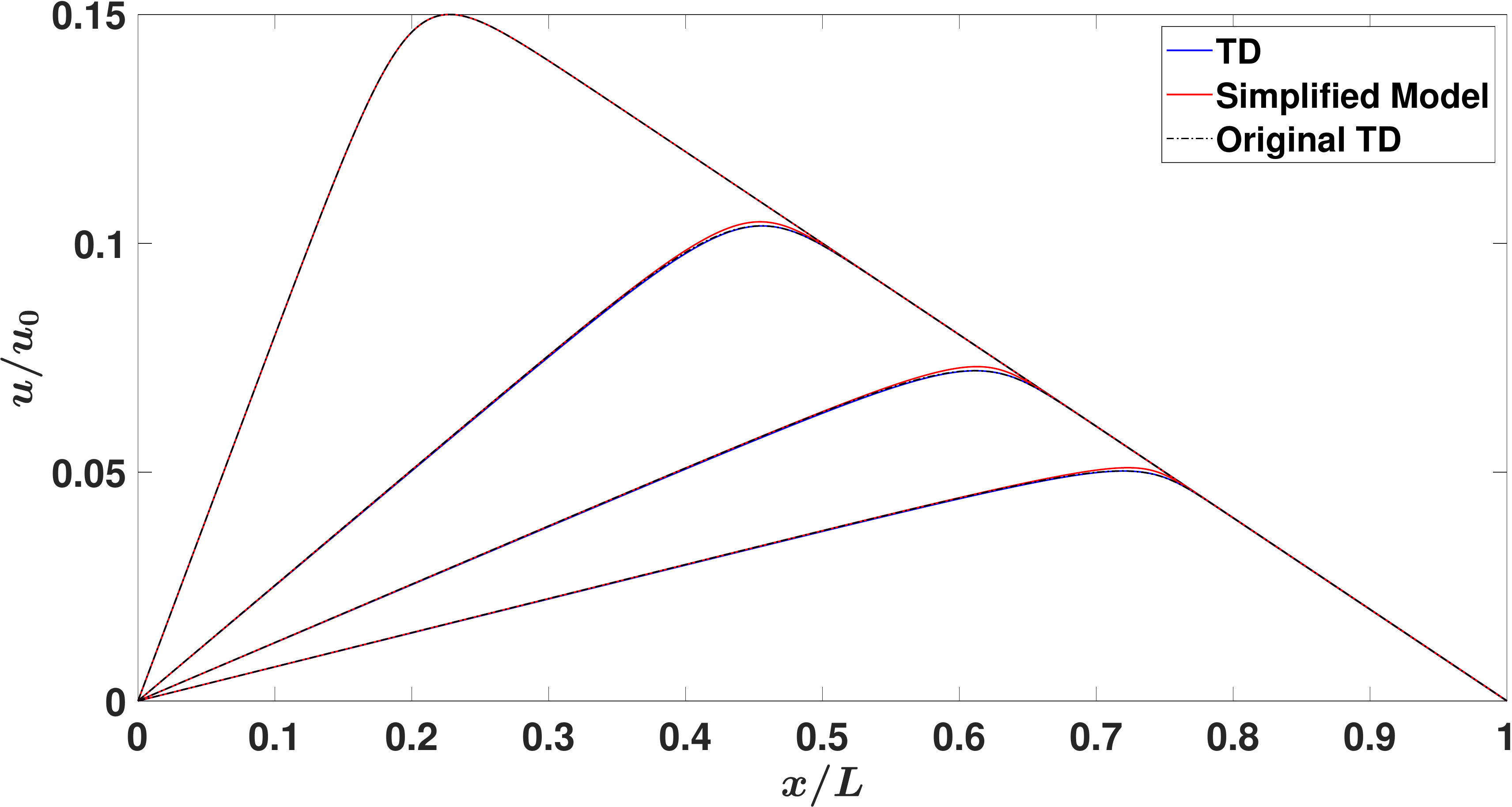}
        \caption{(\textit{b})}
        \label{fig:V_low_M_low_Pe_original}
    \end{subfigure}
    \begin{subfigure}[t]{0.5\textwidth}
        \centering
        \includegraphics[width = \textwidth]{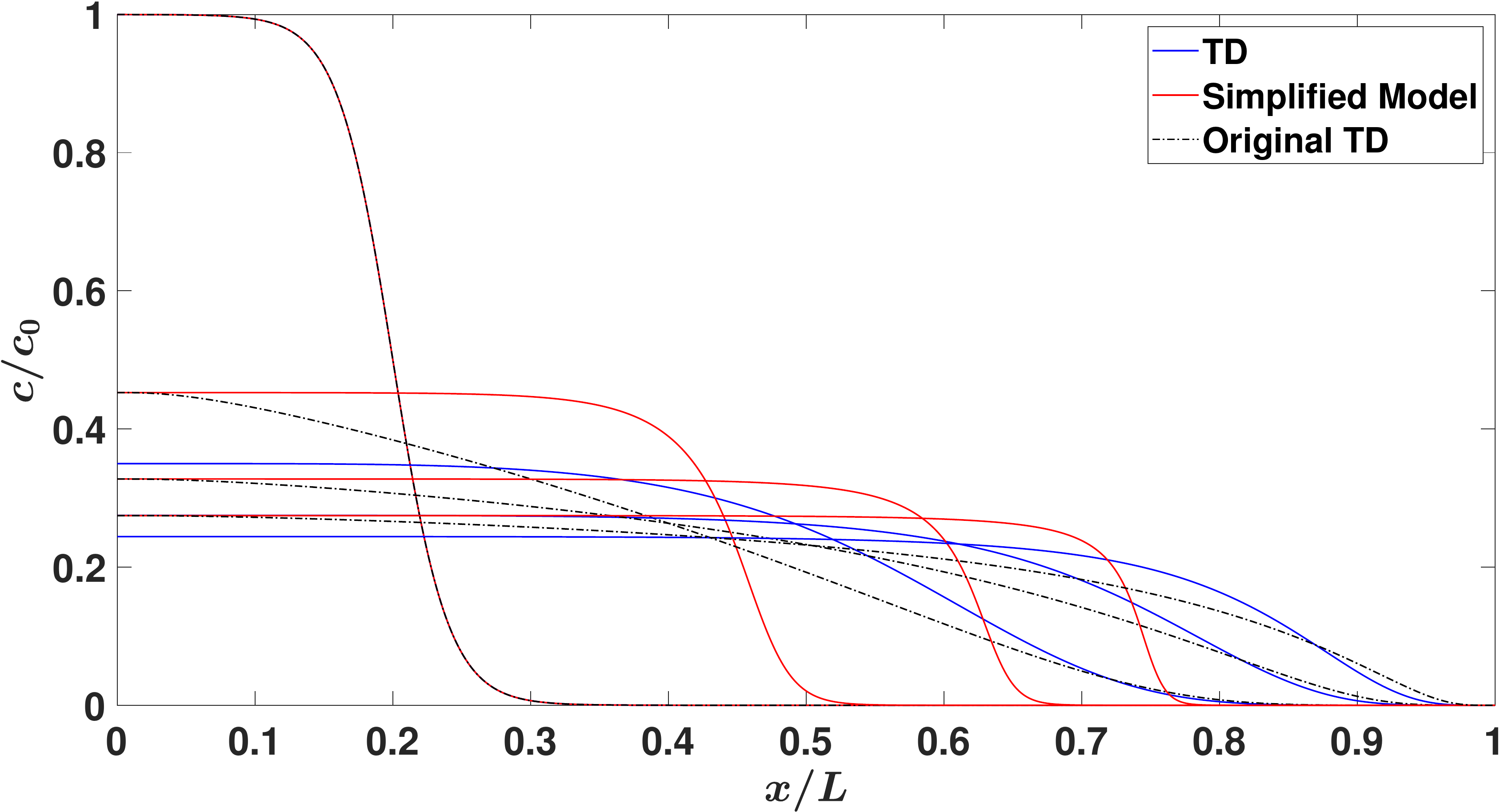}
        \caption{(\textit{c})}
        \label{fig:C_low_M_high_Pe_original}
    \end{subfigure}
    \begin{subfigure}[t]{0.5\textwidth}
        \centering
        \includegraphics[width = \textwidth]{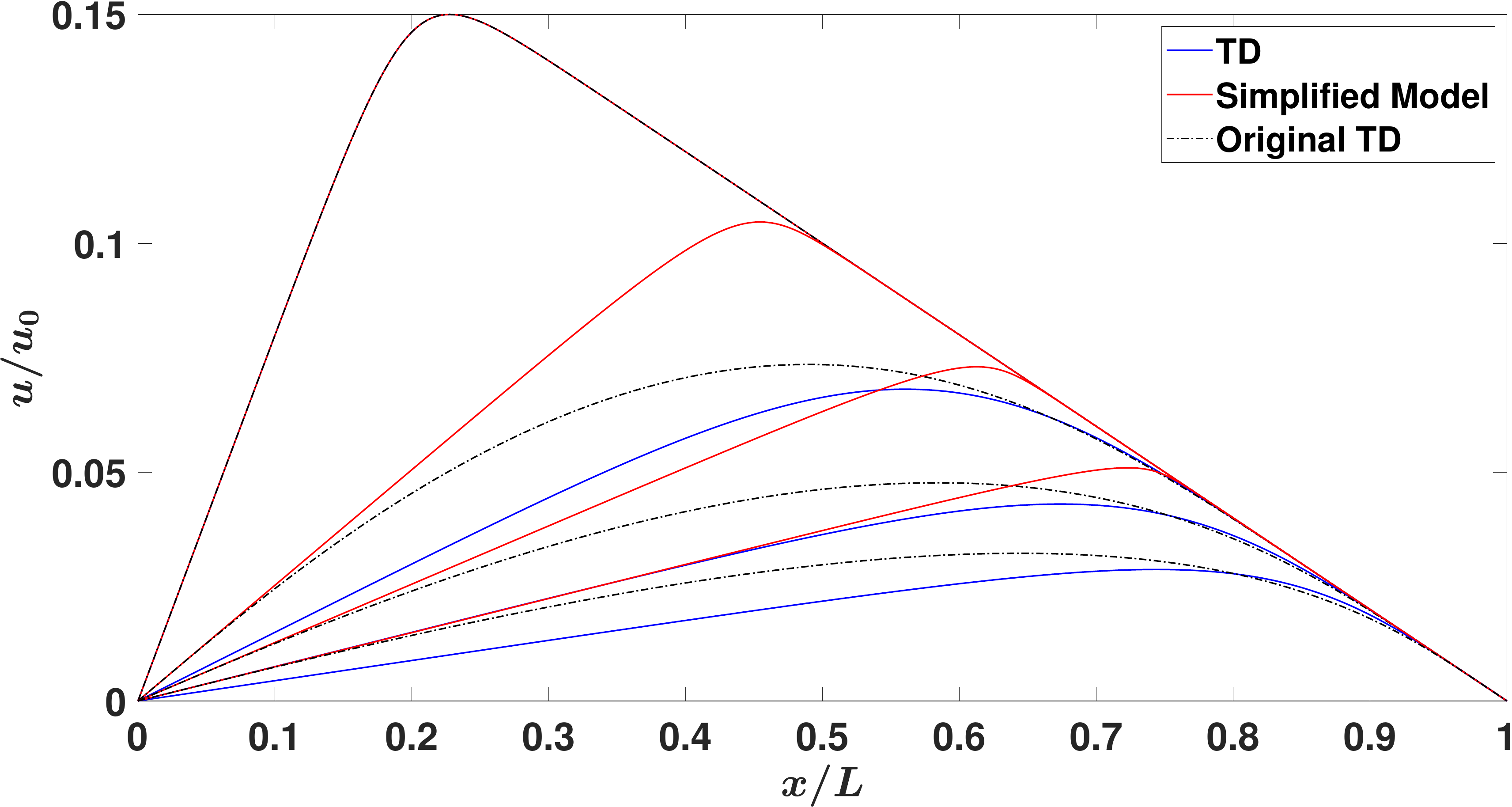}
        \caption{(\textit{d})}
        \label{fig:V_low_M_high_Pe_original}
    \end{subfigure}
    \caption{Results of the numerical solution for the first and second case in table \ref{tab:cases} (low $M$). (\textit{a}) time evolution of the concentration and (\textit{b}) time evolution of the longitudinal mean velocity along the tube length for low $\Pen_r$. (\textit{c}) time evolution of the concentration and (\textit{d}) time evolution of the longitudinal mean velocity along the tube length for high $\Pen_r$. Solid blue lines denote the result with Taylor dispersion (TD), solid red lines denote the result without Taylor dispersion (simplified model) and dashed-dotted (-.) denote the result with original Taylor dispersion (i.e. accounting for $D_d$ only).}
    \label{fig:low_M_original}
\end{figure}

In the second regime, re-scaling by $M$ leads to
\begin{equation}
    \begin{split}
        \frac{1}{M}\frac{\p U}{\p X} = C - P,\\
        U = \frac{\p P}{\p X},
        \label{eq:naming_nondimensional_high_M}
    \end{split}
\end{equation}
where $p_0 = R_g T c_0$ and $u_0 = a^2 p_0 (8 \mu L)^{-1} = a^2 R_g T c_0 (8 \mu L)^{-1}$. In this case, the pressure is directly dependent on the osmotic potential and the velocity is scaled from the Navier-Stokes equations. The naming for these two cases was primarily related to the scaling of the pressure that is implicit in equation (\ref{membrane_physics}).

\section{}\label{appC}
This appendix explores the effect of the new advection correction term that arises from the Taylor dispersion analysis in osmotically driven flow. A comparison between the solutions that includes the new advection term (i.e. TD) and the solution that ignores the advection term while including the diffusional effect (resembling in mathematical form the original Taylor dispersion) are discussed.  The model that ignores entirely TD effects is used as a reference (i.e. 'simplified model').
\par
The formulation for the original Taylor dispersion in the absence of the advection term is a limit set by assuming that the radial advection term in equation (\ref{eq:advection_diffusion}) is much smaller than the diffusion term. In this case, equation (\ref{eq:cm_TD}) reduces to
\begin{equation}
  \frac{\p \bar{c}}{\p t} + \frac{\p}{\p x}\left(\bar{c} \bar{u}\right) = \frac{\p}{\p x}\left[\left(\frac{a^2 \bar{u}^2}{48 D} + D\right)\frac{\p \bar{c}}{\p x}\right].
  \label{eq:cm_TD_original}
\end{equation}

\begin{figure}
    \begin{subfigure}[t]{0.5\textwidth}
        \centering
        \includegraphics[width = \textwidth]{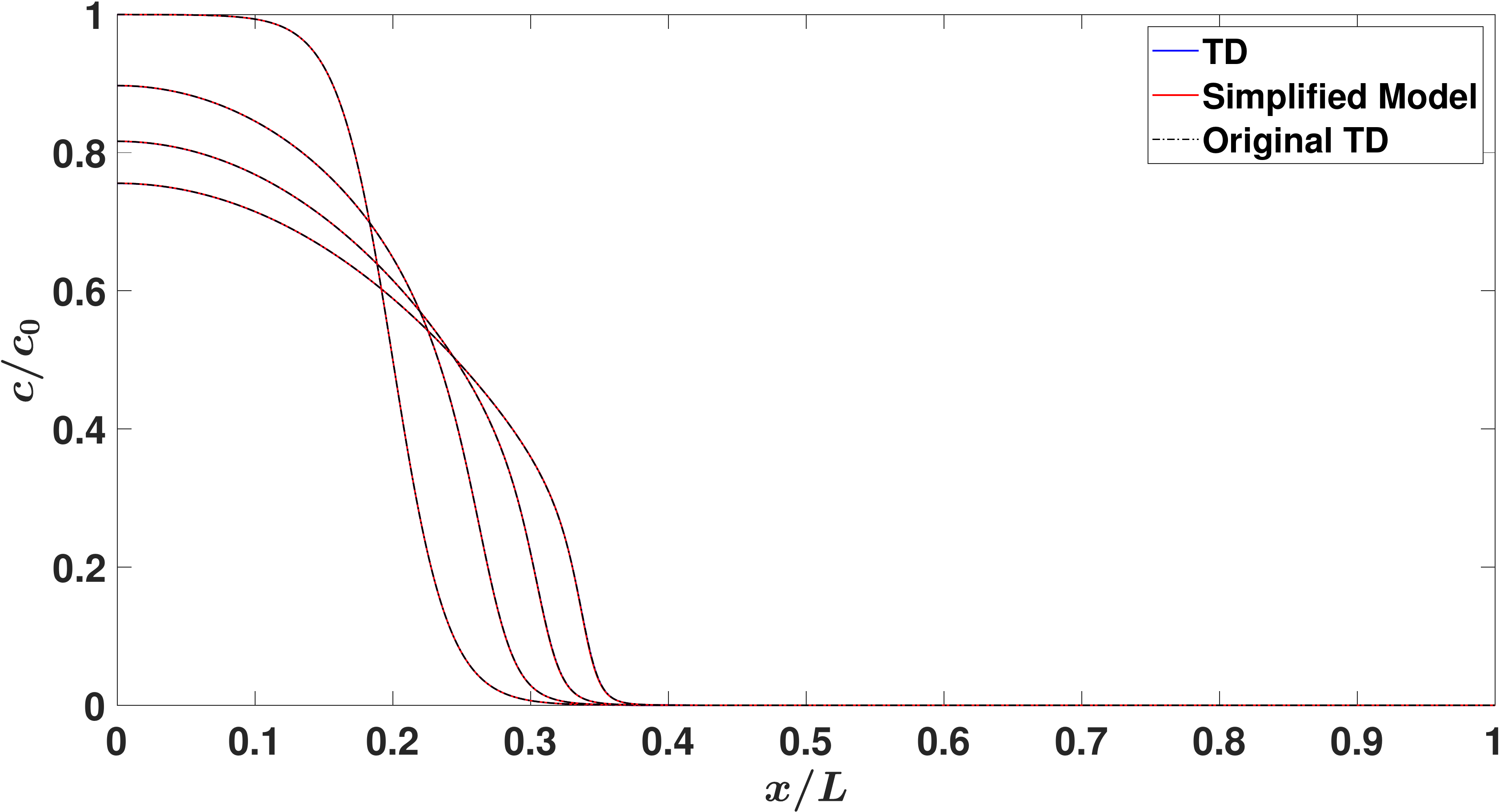}
        \caption{(\textit{a})}
        \label{fig:C_high_M_low_Pe_original_usingV}
    \end{subfigure}
    \begin{subfigure}[t]{0.5\textwidth}
        \centering
        \includegraphics[width = \textwidth]{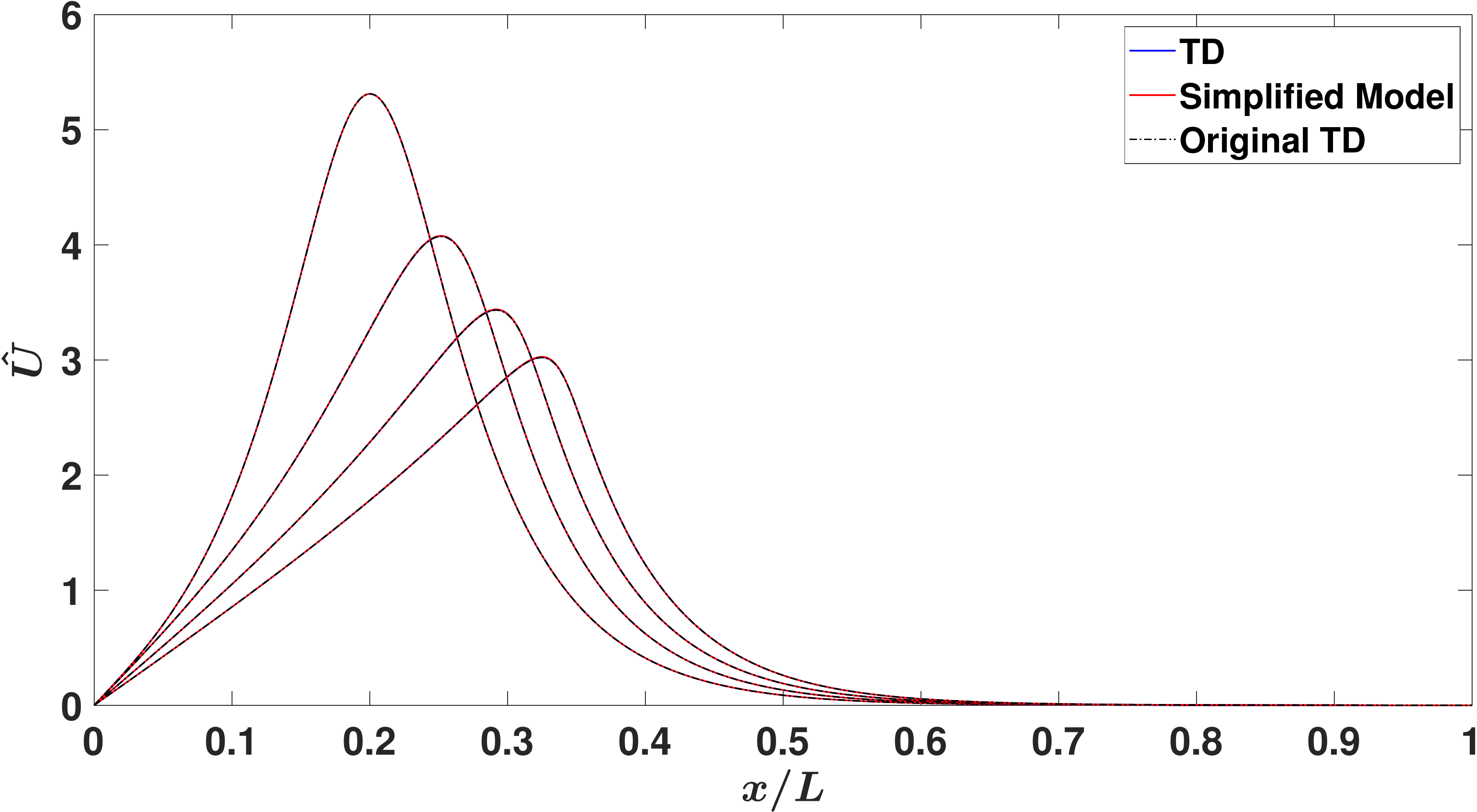}
        \caption{(\textit{b})}
        \label{fig:V_high_M_low_Pe_original_usingV}
    \end{subfigure}
    \begin{subfigure}[t]{0.5\textwidth}
        \centering
        \includegraphics[width = \textwidth]{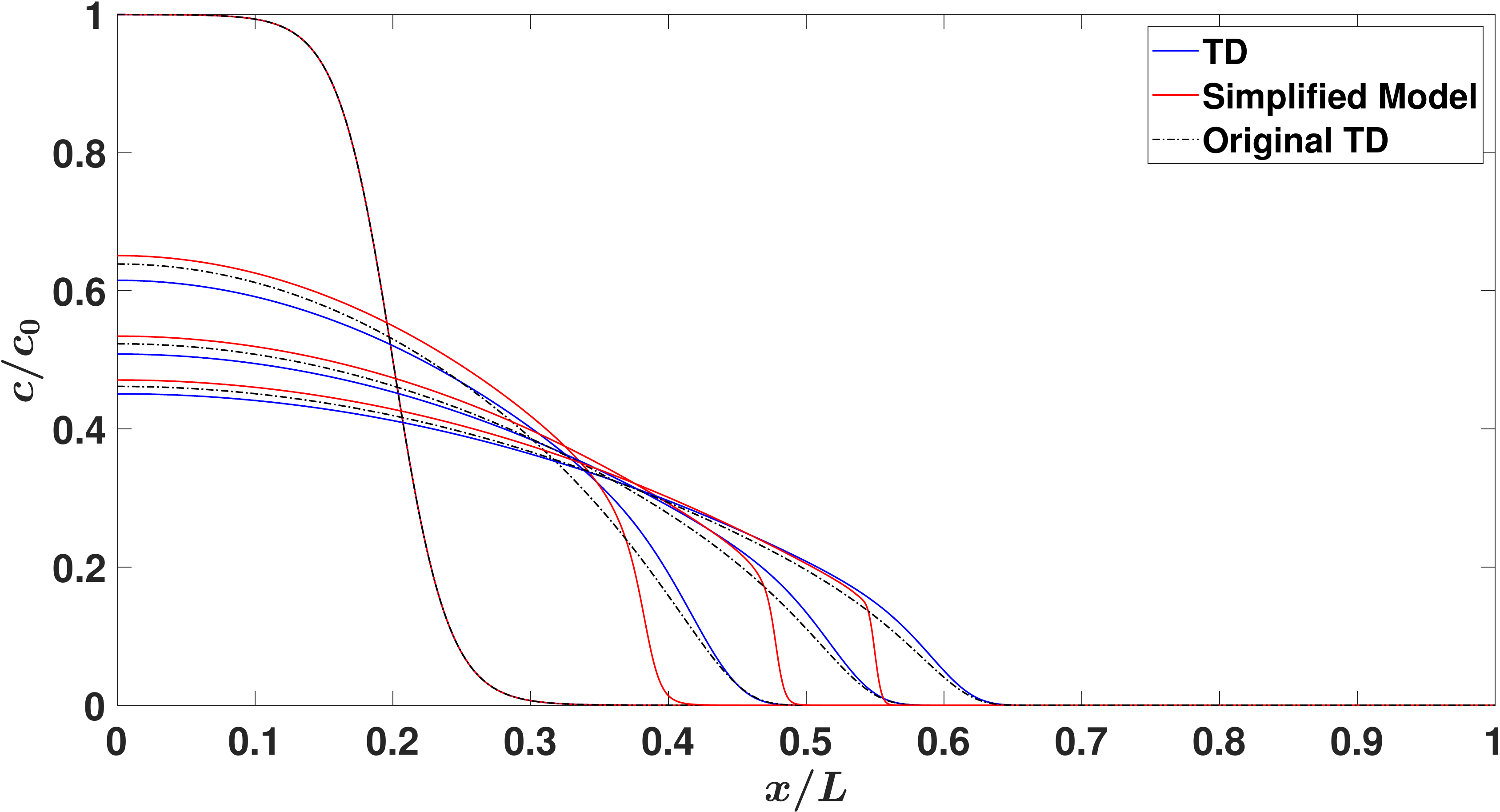}
        \caption{(\textit{c})}
        \label{fig:C_high_M_high_Pe_original_usingV}
    \end{subfigure}
    \begin{subfigure}[t]{0.5\textwidth}
        \centering
        \includegraphics[width = \textwidth]{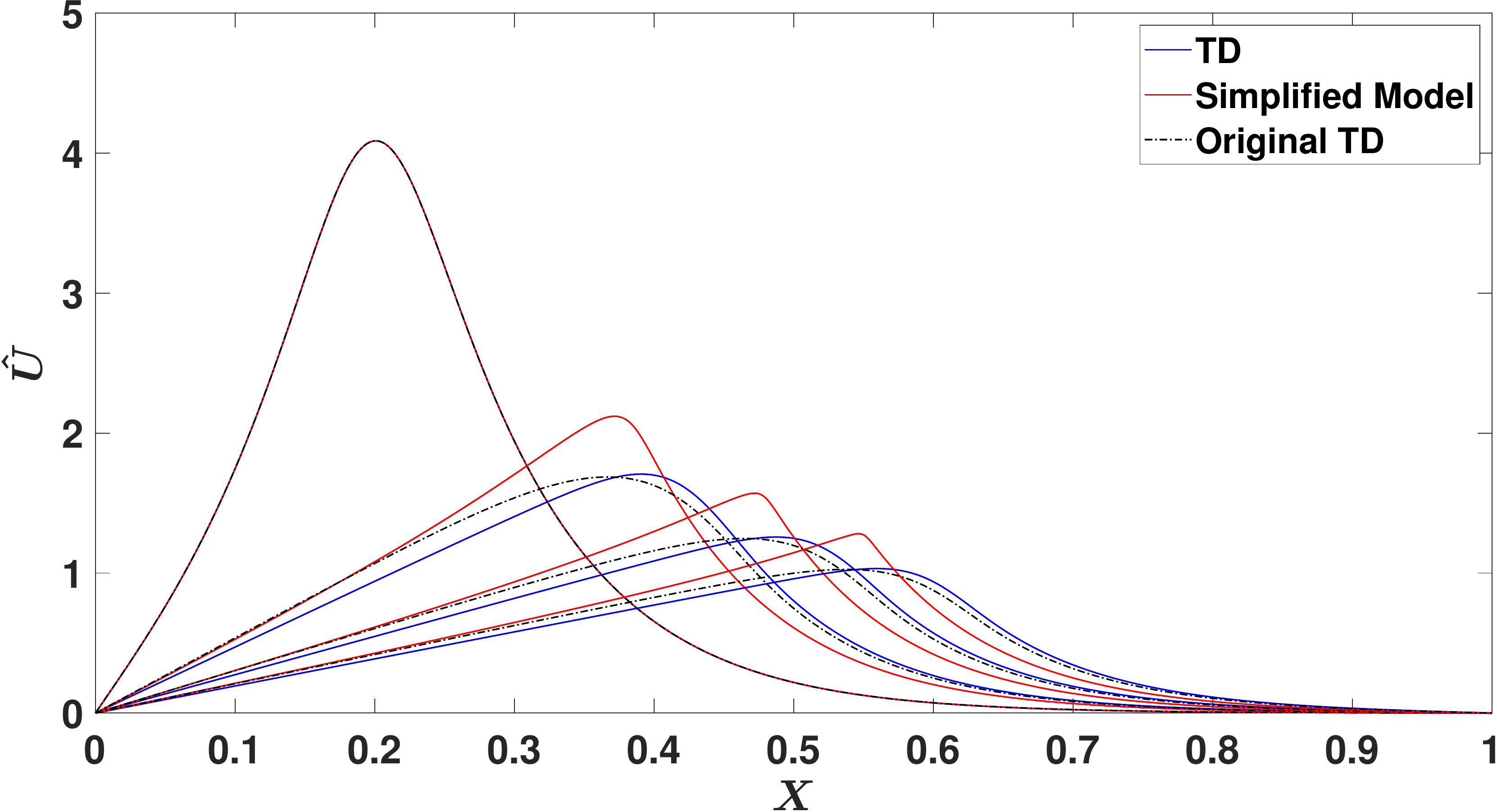}
        \caption{(\textit{d})}
        \label{fig:V_high_M_high_Pe_original_usingV}
    \end{subfigure}
    \caption{Results of the numerical solution for the third and fourth case in table \ref{tab:cases} (high $M$). (\textit{a}) time evolution of the concentration and (\textit{b}) time evolution of the velocity profile $\hat{U}$ for low $\Pen_r$. (\textit{c}) time evolution of the concentration and (\textit{d}) time evolution of the velocity profile $\hat{U}$ for high $\Pen_r$. Solid blue lines denote the result with Taylor dispersion (TD), solid red lines denote the result without Taylor dispersion as before (simplified model) and dashed-dotted (-.) denote the result with original Taylor dispersion.}
    \label{fig:high_M_original_usingV}
\end{figure}

The outcome in equation (\ref{eq:cm_TD_original}) recovers the original Taylor dispersion result in the sense that the local effect of the advection term is ignored knowing that its global effect will not affect the equation (mass is still conserved globally in the 1D approximation). However, the coupling between $\overline{p}$ and $\overline {c}$ is maintained by the membrane physics and the van't Hoff relation $R_g T \overline {c}=\overline{p}$ so that $\overline{u}$ is fully described by equation (\ref{membrane_physics}). Equations (\ref{eq:cm_TD_original}) and (\ref{membrane_physics}) provide a solution for $\overline{u}(x)$ and $\overline{c}(x)$ in the limit where radial advection is negligible and its effect as local sink or source for $\bar{c}$ is ignored.
\par
As expected, the original Taylor dispersion model will not affect the model globally meaning that the speed of the flow is still approximately the same as the one that includes both terms where both models will reach the end of the tube at the same time. However, locally, the effect is apparent especially for the high $\Pen_r$ number cases as shown in figures \ref{fig:C_low_M_high_Pe_original} and \ref{fig:C_high_M_high_Pe_original_usingV}.

\begin{figure}
    \begin{subfigure}[t]{0.5\textwidth}
        \centering
        \includegraphics[width = \textwidth]{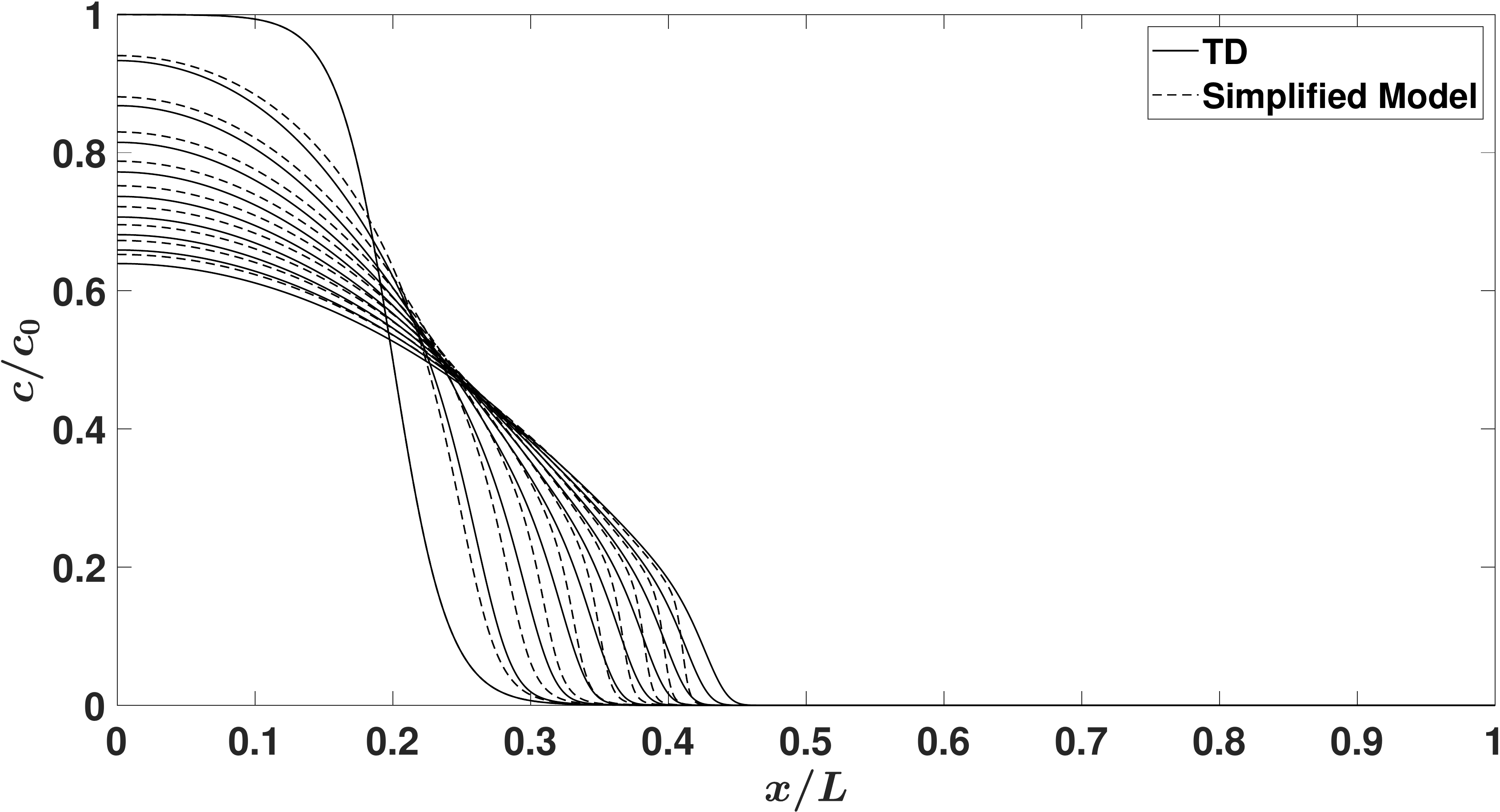}
        \caption{(\textit{a})}
        \label{fig:C_high_M_low_Pe}
    \end{subfigure}
    \begin{subfigure}[t]{0.5\textwidth}
        \centering
        \includegraphics[width = \textwidth]{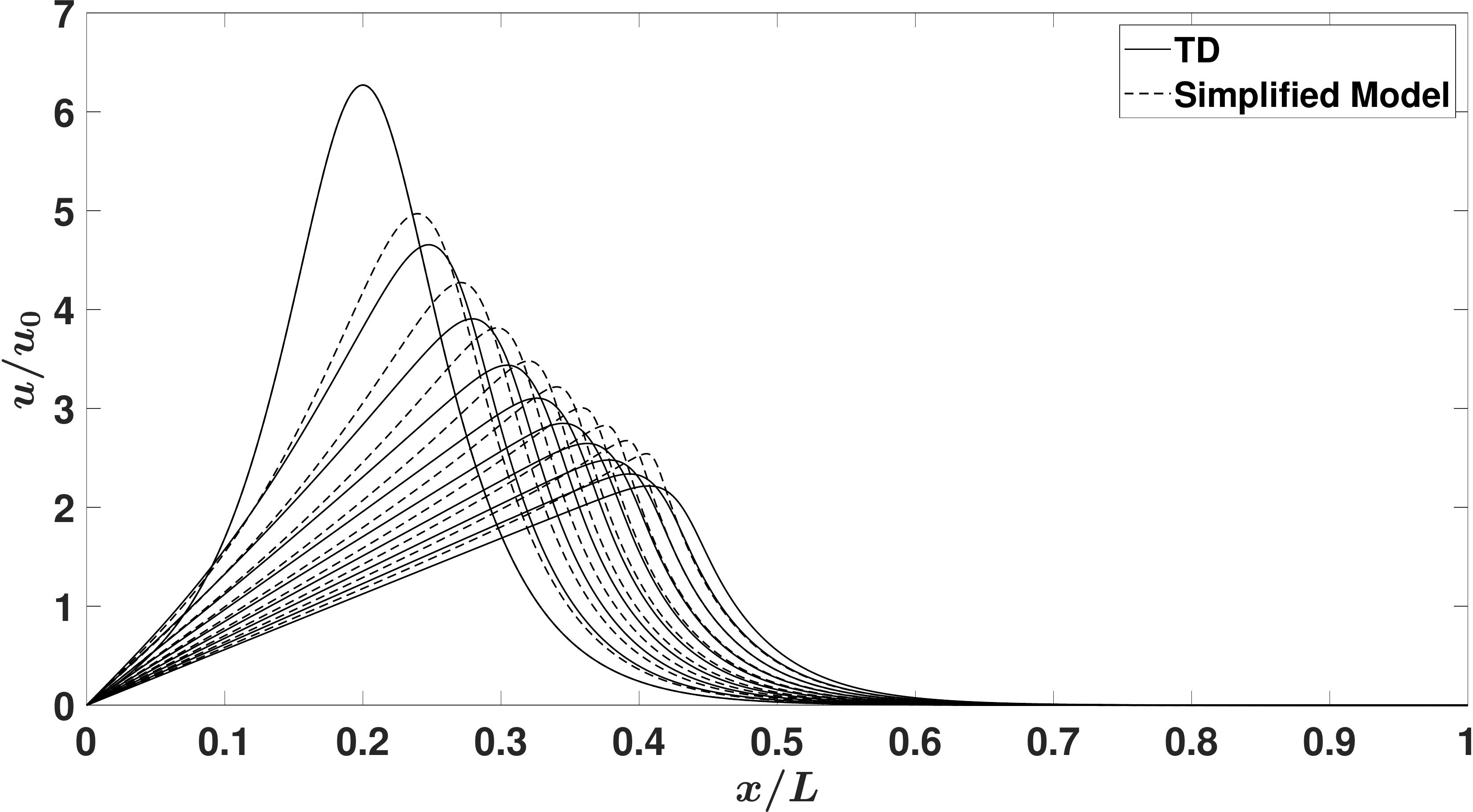}
        \caption{(\textit{b})}
        \label{fig:V_high_M_low_Pe}
    \end{subfigure}
    \begin{subfigure}[t]{0.5\textwidth}
        \centering
        \includegraphics[width = \textwidth]{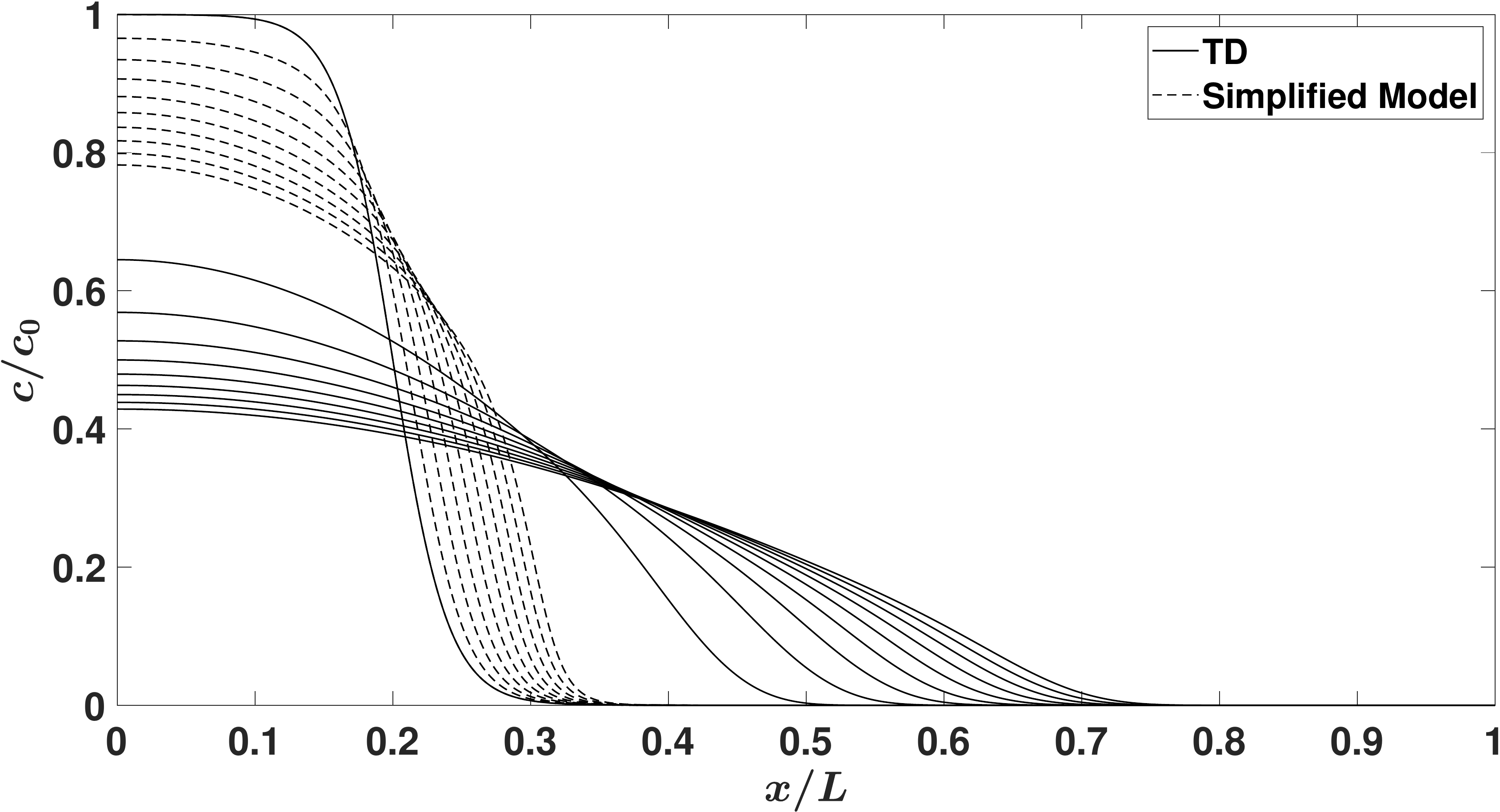}
        \caption{(\textit{c})}
        \label{fig:C_high_M_high_Pe}
    \end{subfigure}
    \begin{subfigure}[t]{0.5\textwidth}
        \centering
        \includegraphics[width = \textwidth]{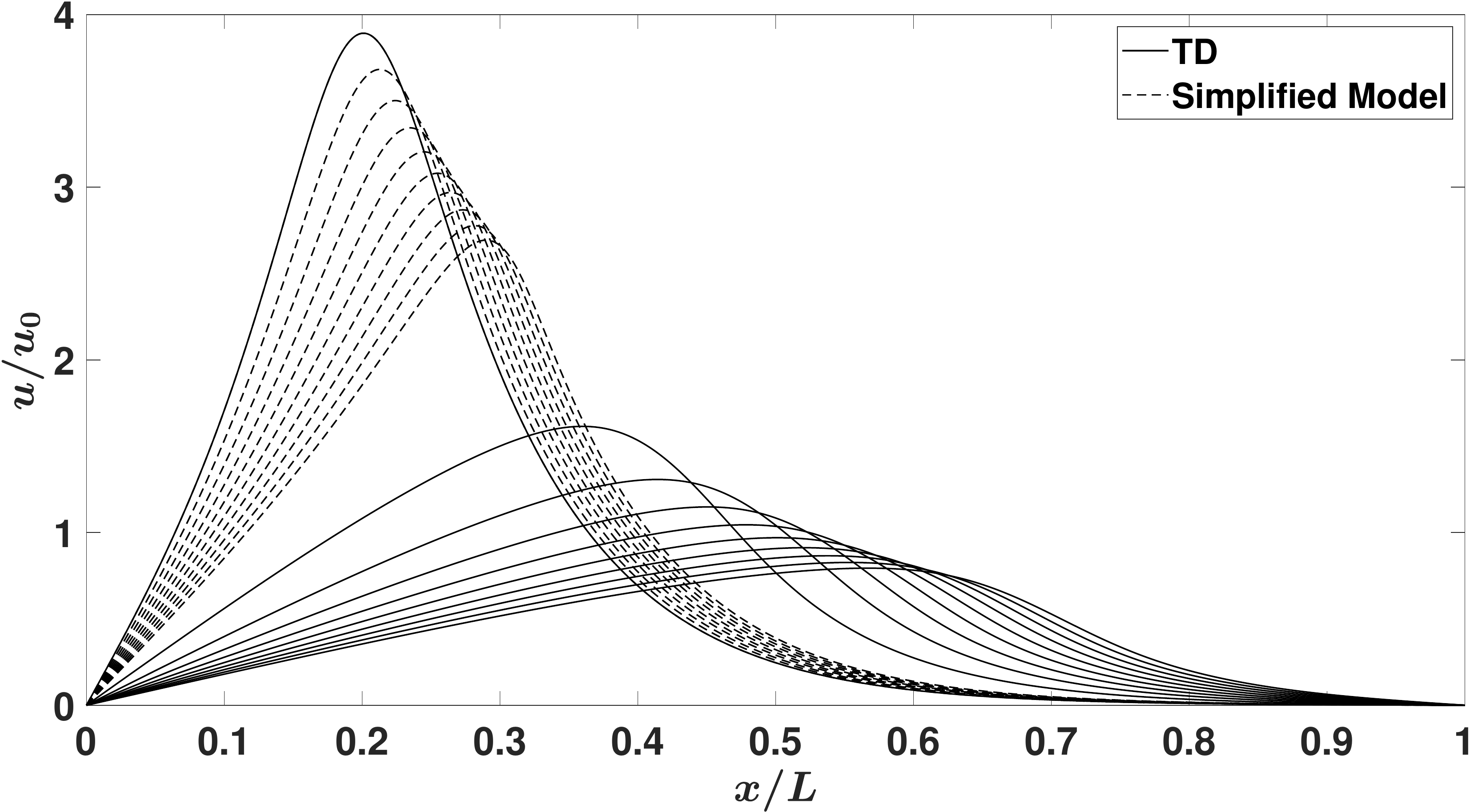}
        \caption{(\textit{d})}
        \label{fig:V_high_M_high_Pe}
    \end{subfigure}
    \caption{Results of the numerical solution for high $M$ (third and fourth case in table \ref{tab:cases}). (\textit{a}) time evolution of the concentration and (\textit{b}) time evolution of the velocity profile for low $\widehat{\Pen_r}$. (\textit{c}) time evolution of the concentration and (\textit{d}) time evolution of the velocity profile for high $\widehat{\Pen_r}$. Solid lines denote the result with Taylor dispersion and dashed lines denote the result without Taylor dispersion as before.}
    \label{fig:high_M}
\end{figure}

\par
For small $M$, the original Taylor dispersion model is more diffusive because the advection term that can locally behave as source or sink is ignored. It is apparent from these figures that near $X = 0$, the removal of the concentration in the full Taylor dispersion model is much faster because in this region $du/dx$ is positive. However, at the end of the tube, both models have the same speed because in this region the gradient is negative and will slow down the front speed.
\par
For large $M$, two differences can arise from neglecting the advection term when $\Pen_r > 1$. First, the speed to attain a self similar solution increases by invoking each step of approximation: fastest for the full TD, slower for original TD and slowest for the model that neglects both. Second, the analytical solution in the moving boundary layer is now different for each model.

\section{}\label{appD}
In this appendix, the work will focus on showing how $\Pen_r$ can have a higher effect on the flow for the large $M$ number regime, while making sure that inertial forces in equation (\ref{eq:NSE_nondimensional}) can still be neglected. As discussed in section \ref{sec:nondimensional_simplified_model}, in the large $M$ case, the axial velocity scales as $1/M \sim \epsilon_1$. This means that the dimensional velocity is very small and the radial P\'eclet number scales as $\Pen_r \sim \textit{O}(10) M^{-1}$. In this case, one can get a new velocity scale, in order to make the axial velocity scales as $\sim \textit{O}(1)$. This scale can be achieved by working on the second case in appendix \ref{appB} (i.e. for large $M$), or by taking equation (\ref{eq:membrane_physics}) and using the following scales for the velocity, concentration and time: $u = \widehat{u_0} U$, $c = c_0 C$ and $t = \widehat{t_0} \tau$. In this case, the nondimensional form of equation (\ref{eq:membrane_physics}) can be written as:
\begin{equation}
    \frac{\p C}{\p X} = \frac{1}{M} \frac{\p^2 U}{\p X^2} - U.
\end{equation}
where $U = \textit{O}(1)$ and $C = \textit{O}(1)$. In this case, the nondimensional form of equation (\ref{eq:cm_TD}) is the same as equation (\ref{eq:cm_TD_nondimensional_finite_M})
\begin{equation}
    \frac{\p C}{\p \tau} + \frac{\p}{\p X}\left[\left(1 + \frac{\widehat{\Pen_r}}{24}\frac{\p U}{\p X}\right)C U\right] = \frac{\widehat{\Pen_r}}{48} \frac{\p}{\p X}\left[\left(U^2 + \frac{48}{\widehat{\Pen_r} \widehat{\Pen_l}}\right)\frac{\p C}{\p X}\right],
    \label{eq:cm_TD_nondimensional_high_M_rescaled}
\end{equation}
where $\widehat{u_0} = u_0/M$, $\widehat{\Pen_r} = \Pen_r/M$, $\widehat{\Pen_l} = \Pen_l/M$ and $\widehat{t_0} = t_0 M$.
For this type of flow, the largest order of magnitude that $\widehat{\Pen_r}$ can achieve is also $\textit{O}(10)$ since both $\epsilon \Rey$ and $\Pen_r$ were re-scaled by $M$ and their ratio is still the same (i.e. $\mu (\rho D)^{-1}$).
\par
First, starting with the $\widehat{\Pen_r} \ll 1$ case, the following set of variables were chosen:\\
$k = 5 \times 10^{-11}$ m (Pa s)$^{-1}$, \qquad $L = 3$ m, \qquad $a = 3 \times 10^{-5}$ m, \qquad $c_0 = 100$ mMol\\
Figures \ref{fig:C_high_M_low_Pe} and \ref{fig:V_high_M_low_Pe} show the result for the concentration and velocity profiles respectively. As one can see from these figures, the results are very similar to the case where $\Pen_r \sim \textit{O}(10)$ in section \ref{sec:TD_nondimensional_osmotically}. It should be noted here that a linear van't Hoff relation between the osmotic pressure and the concentration is used even though the value for $c_0$ is higher than the one used in section \ref{sec:TD_nondimensional_viscous} for simplicity.
\par
As in section \ref{sec:TD_nondimensional_viscous}, the interesting case is the higher $\widehat{\Pen_r}$. For this reason, the set of initial and geometrical conditions chosen for illustration are:\\
$k = 6 \times 10^{-10}$ m (Pa s)$^{-1}$, \qquad $L = 5$ m, \qquad $a = 1.5 \times 10^{-4}$ m, \qquad $c_0 = 200$ mMol\\
Figures \ref{fig:C_high_M_high_Pe} and \ref{fig:V_high_M_high_Pe} reveal a different self similar solution for the concentration distribution for the Taylor dispersion model than the model that ignores Taylor dispersion as shown in figure \ref{fig:C_high_M_high_Pe}. Re-scaling equation (\ref{eq:cm_TD_nondimensional_high_M_rescaled}) by $\widehat{\Pen_r}/24$ as before and using a linear relation between velocity and the concentration, the nondimensional form for this type of flow can be expressed as 
\begin{equation}
  \frac{\p C}{\p \tau} = \frac{\p}{\p X}\left[\left[\left(\frac{24}{\widehat{\Pen_r}} - \frac{\p^2 C}{\p X^2}\right) C + \frac{1}{2}\left(\frac{\p C}{\p X}\right)^2\right]\frac{\p C}{\p X}\right],
  \label{eq:cm_TD_nondimensional_osmotically_large_Pe_rescaled}
\end{equation}
where the time scale follows the same derivation as in section \ref{sec:TD_nondimensional_viscous} for $\Pen_r/24 > 1$ and the molecular diffusion term has been neglected.
\par
In this case, if equation (\ref{eq:cm_TD_nondimensional_osmotically_large_Pe_rescaled}) is expanded as an infinite series of different orders for the concentration, a self similar solution for the first order term emerges. However, this analysis is also better kept for a future inquiry.
\bibliographystyle{jfm}
\bibliography{sugar_transport_references}

\end{document}